%Paper: astro-ph/9404062
%From: dong@astrosun.tn.cornell.edu (Dong Lai)
%Date: Sun, 24 Apr 1994 21:31:08 -0400

%\magnification=1200
\baselineskip=18 pt
\font\bigbf=cmbx10 scaled 1400

\def\hi{\noindent \hangindent=2.5em}
\def\go{\mathrel{\raise.3ex\hbox{$>$}\mkern-14mu
             \lower0.6ex\hbox{$\sim$}}}
\def\lo{\mathrel{\raise.3ex\hbox{$<$}\mkern-14mu
             \lower0.6ex\hbox{$\sim$}}}
\def\al{{\alpha}}
\def\eps{{\epsilon}}
\def\br{{\bf r}}
\def\be{{\bf e}}
\def\bD{{\bf D}}
\def\bxi{{\vec \xi}}
\def\xir{{\xi^r}}
\def\xip{{\xi^\perp}}
\def\cN{{\cal N}}

\bigskip
\bigskip
%%%%%%%%%%%%%%%%%%%%%%%%%%%%%%%%%%%%%%%%%%%%%%%%%%%%%%%%%%%%%%%%%%%%%%
\centerline{\bigbf RESONANT OSCILLATIONS AND TIDAL HEATING}
\medskip
\centerline{\bigbf IN COALESCING BINARY NEUTRON STARS}
\bigskip

\centerline{
DONG LAI$\,$\footnote{$^1$}
{Department of Physics, Cornell University.
E-mail: dong@astrosun.tn.cornell.edu}
}
\medskip
\centerline{\it Center for Radiophysics and Space Research,
Cornell University}
\smallskip
\centerline{\it Ithaca, NY 14853, USA}

\vskip 1.0truecm
%%%%%%%%%%%%%%%%%%%%%%%%%%%%%%%%%%%%%%%%%%%%%%%%%%%%%%%%%%%%%%%%%%%%%%%%
\centerline{\bf ABSTRACT}
\vskip 0.3truecm

Tidal interaction in a coalescing neutron star binary can resonantly excite
the g-mode oscillations of the neutron star when the frequency of the tidal
driving force equals the intrinsic g-mode frequencies. We study the g-mode
oscillations of cold neutron stars using recent microscopic nuclear equations
of state, where we determine self-consistently the sound speed and
Brunt-V\"ais\"al\"a frequency in the nuclear liquid core. The properties of
the g-modes associated with the stable stratification of the core depend
sensitively on the pressure-density relation as well as the symmetry energy
of the dense nuclear matter. The frequencies of the first ten g-modes lie
approximately in the range of $10-100$ Hz. Resonant excitations of these
g-modes during the last few minutes of the binary coalescence result in energy
transfer and angular momentum transfer from the binary orbit to the neutron
star. The angular momentum transfer is possible because a dynamical tidal lag
develops even in the absence of fluid viscosity. However, since the coupling
between the g-mode and the tidal potential is rather weak, the amount of
energy transfer during a resonance and the induced orbital phase error are
very small.

Resonant excitations of the g-modes play an important role in tidal heating
of binary neutron stars. Without the resonances, viscous dissipation is
effective only when the stars are close to contact. The resonant oscillations
result in dissipation at much larger orbital separation. The actual amount of
tidal heating depends on the viscosity of the neutron star. Using the
microscopic viscosity, we find that the binary neutron stars are heated
to a temperature $\sim 10^8$ K before they come into contact.

\vskip 0.5truecm
{\bf Key Words:} binaries: close --- radiation mechanisms: gravitational
--- stars: neutron --- oscillations --- hydrodynamics --- equation of state
--- dense matter

\bigskip
\bigskip
\centerline{\it Submitted to MNRAS, December, 1993}
\medskip
\centerline{\it Accepted April, 1994}

\vfil\eject
%\bigskip
%%%%%%%%%%%%%%%%%%%%%%%%%%%%%%%%%%%%%%%%%%%%%%%%%%%%%%%%%%%%%%%%%%%%%%%%%%
\centerline{\bf 1. INTRODUCTION}
\medskip

Coalescing neutron star-neutron star (NS-NS) and neutron star-black hole
(NS-BH) binaries are believed to be one of
the most promising sources of gravitational radiation (e.g., Schutz 1986,
Thorne 1987) that could be detected by
the new generation of instruments such as the Laser
Interferometer Gravitational-Wave Observatory (LIGO, see Abramovici et al 1992)
and its French-Italian counterpart VIRGO (Bradaschia et al 1990).
Extrapolation of the estimated galactic formation rates of compact
binaries based on the observed local population of
binary radio pulsars leads to an estimate of the rate of coalescence
in the Universe of about $10^{-7}$ yr$^{-1}$ Mpc$^{-3}$
(Narayan, Piran \& shemi 1991, Phinney 1991).
Extracting gravity wave signals from noise requires accurate theoretical
waveforms in the frequency range $10-1000$ Hz, corresponding to the last few
minutes of the binaries' life (e.g, Cutler et al. 1993). Effects
of tidal interaction on the orbital decay and gravitational radiation waveforms
have been found to be small except during the final stage of coalescence
(Kochanek 1992, Bildsten \& Cutler 1992,
Lai, Rasio \& Shapiro 1993,~1994a,~1994b).
However, all these studies implicitly assume equilibrium or quasi-equilibrium
tides, while dynamical tidal responses of the stars are completely
neglected. Such dynamical tides are possible when the interaction potential
between the two components of the binary resonantly excites the
intrinsic oscillations of the neutron stars.

Recent studies of neutron star oscillations have
demonstrated the existence of low frequency g-modes of various kinds.
McDermott, Van Horn \& Scholl (1983) and McDermott, Van Horn \& Hansen (1988)
calculated the g-modes associated with thermally induced buoyancy for warm
(hot) neutron stars. Finn (1987) considered crustal g-modes due to
the composition discontinuities in the outer envelope of cold neutron stars.
Reisenegger \& Goldreich (1992, hereafter RG1) studied g-modes associated
with the buoyancy induced by proton-neutron composition gradient in the cores
of neutron stars. Regardless of the nature of the g-mode oscillations and
various theoretical uncertainties, the existence of such g-modes indicates
that a neutron star in a binary system can be resonantly excited when the
frequency of the tidal driving force equals the g-mode frequencies.
These resonant oscillations and their consequences to the
binary orbital decay and the neutron stars themselves are the subjects of
this paper.

For a star in a binary system, the tidal potential due to the
companion acts like an external perturbing force, with a driving
frequency $2\Delta\Omega=2(\Omega-\Omega_s)$, where
$\Omega$ and $\Omega_s$ are the orbital angular frequency and the spin
angular frequency of the star (the factor of $2$ results from the quadrupole
nature of the tidal potential, and will become
apparent in \S 2). The star behaves as a collection of
harmonic oscillators with different eigen-frequencies. Let us consider
just one such oscillation mode
with frequency $\omega$. Schematically, the
equation governing the tidal distortion $\xi$ can be written as
$$\ddot\xi+\omega^2\xi-{1\over t_{visc}}\dot\xi \propto e^{2i\Delta\Omega t},
\eqno(1.1)$$
where $t_{visc}$ is the timescale for viscous dissipation (or other
dissipations)
in the star. For stationary state, we have
$$\xi\propto {1\over \omega^2-4\Delta\Omega^2-2i\Delta\Omega/t_{visc}}
e^{2i\Delta\Omega t}.\eqno(1.2)$$
Since in general the viscous dissipation timescale is much
larger than the dynamical
timescale of the star, we see that when $2\Delta\Omega=\omega$,
the amplitude of $\xi$ can be very large.

Such resonant oscillations in close binaries were first considered by
Cowling (1941) in his seminal paper on nonradial stellar oscillations.
He noted that the coupling between the tidal potential and the high-order
g-mode is weak, therefore a close resonance is needed to greatly enhance
the equilibrium tide (see also Zahn 1970).
He also suggested that such close resonance will
be destroyed by second order effects due to the large horizontal
displacement of the high-order g-mode. Clearly, for
long-lived binary systems, such as binaries consisting of main sequence stars,
what exactly happens to the stars at resonance is a complicated issue
(e.g., Zahn 1977, Nicholson 1978, Goldreich \& Nicholson 1989).

For an inspiraling neutron star binary, the problem becomes ``cleaner''.
The rapid decay of the orbit induces a dynamical tidal lag, and implies
that the orbital frequency
does not stay at a particular resonant value very long.
Thus the rapid decay of the orbit ``kills'' the infinity
inherent in equation (1.2) even without dissipative or other higher-order
effects. Moreover, we shall see that the resonant tidal distortion is small and
a linear approximation for the excited oscillations is adequate.

In this paper, we shall concentrate on the excitations of core g-modes.
These modes involve the bulk region of a neutron star,
while the g-modes driven by entropy gradient and density discontinuities
are confined to its outer envelope (Finn 1987, McDermott, Van Horn \& Hansen
1988). Previous study (RG1) has yielded a qualitative understanding
of the basic properties of the core g-modes. However, the detailed
properties of these g-modes, especially the strengths of their coupling to
the tidal field, which directly affect the amplitudes of the excitations,
depend sensitively on the values of the Brunt-V\"ais\"al\"a frequency in
different regions of the star. The calculations by RG1 were based on
an approximate, inconsistent
ansatz for the Brunt-V\"ais\"al\"a frequency
in the core. We examine some recent microscopic equations of state of nuclear
matter, and use them to obtain the self-consistent
Brunt-V\"ais\"al\"a frequency, upon which our calculations of g-modes
are based (see \S 4 and \S 5).

A related question of interest concerns how much energy is dissipated
into heat in coalescing binary neutron stars. M\'esz\'aros \& Rees (1992)
considered the consequence of such tidal heating in
the context of cosmological Gamma ray burst models.
Significant tidal heating of the neutron stars
can induce mass loss from the stars even before merging takes place.
Such mass loss can lead to baryon contamination of the possible Gamma ray
burst emission. Clearly, this result depends on the viscosity of the neutron
star matter. We show that the resonant excitations of g-modes are important in
the viscous heating of binary neutron stars (\S 8).
%can be important than the
%standard estimate of viscous dissipation via quasi-equilibrium tidal
%interaction (\S 8).

Some aspects of the problem of resonant excitations of g-modes
have also recently been considered by Reisenegger \& Goldreich
(1994, hereafter RG2). They calculated the total energy transfer from the
binary orbit to the star in a resonance
using the formalism of Press \& Teukolsky (1977, hereafter PT), who considered
a similar problem in the context of parabolic encounter of two stars.
%and the resulting tidal capture.
%Strictly speaking, the PT formalism does not exactly apply to the
%situation considered here: while the parabolic fly-by is a process
%occurring from $t=-\infty$ to $t=\infty$, the binary inspiraling terminates
%when the star comes into contact. Also, to study the time-dependence of
%energy transfer process (rather than the total energy transferred),
%the PT formalism is not adequate.
We present here a different, more complete analysis of oscillations
in coalescing binary neutron stars, including
dynamical tidal lag, energy transfer and angular momentum transfer
due to resonances, etc. Our method enables us to study the
time-dependence of the resonant energy
transfer process (rather than the total energy transferred),
as well as the non-resonant, quasi-equilibrium tidal effects.
We include comparisons of our results with those of RG2.

In \S 2 we develop the formalism to study tidal excitations of normal modes
in inspiraling binary stars, where we show that the
problem can naturally be divided into two parts. The dynamical part is
discussed in \S 3, while the physics of core g-modes and the calculations
of the mode properties are discussed in \S 4 and \S 5. In \S 6 energy
transfer and angular momentum transfer from the binary orbit to the star
are considered. The effects of such transfer on the
orbital decay and the induced orbital phase error
are discussed in \S 7. Tidal heating of binary neutron stars are considered
in \S 8. We present our conclusions in \S 9, where we also discuss
the importance of g-mode resonances in coalescing white dwarf binaries.

%%%%%%%%%%%%%%%%%%%%%%%%%%%%%%%%%%%%%%%%%%%%%%%%%%%%%%%%%%%%%%%%%%%%%%
\bigskip
%\vfil\eject
\centerline{\bf 2. DYNAMICAL TIDAL INTERACTION IN COALESCING BINARIES:}
\smallskip
\centerline{\bf THE FORMALISM}
\medskip

Consider a star with mass $M$ and radius $R$
in circular orbit with its companion, which we treat as a point
object with mass $M'$. When $M'$ also has a finite size, there is
additional quadrupole-quadrupole coupling between $M$ and $M'$;
but this is a higher-order effect which will be neglected.
Also, general relativistic effects introduce a small correction to
our analysis; these effects are neglected.

The interaction potential due to $M'$ is
$$U=-{GM'\over |\br-\bD|},\eqno(2.1)$$
where $\br$ is the position vector of a fluid element in star M,
and $\bD=\bD(t)$ specifies the position of the point mass. In a comoving
coordinate system $(r,\theta,\phi)$ with its origin at the center of $M$
and the orbital plane at $\theta=\pi/2$, we have
$\bD(t)=(D(t),\pi/2,\Phi(t))$, and the potential can be
expanded into spherical harmonics
$$\eqalign{
U &=-GM'\sum_{lm}{4\pi\over 2l+1}{r^l\over D^{l+1}}Y_{lm}^\ast
\left({\pi\over 2},\Phi\right)Y_{lm}(\theta,\phi)\cr
&= -GM'\sum_{lm}W_{lm}{r^l\over D(t)^{l+1}}
e^{-im\Phi(t)}Y_{lm}(\theta,\phi),\cr
}\eqno(2.2)$$
(using the notation of PT), where the numerical coefficient $W_{lm}$ is
$$W_{lm}
%=\left[{4\pi\over 2l+1}{(l-m)!\over (l+m)!}\right]^{1/2}P_l^m (\cos\Theta=0).
=(-)^{(l+m)/2}\left[{4\pi\over 2l+1}(l+m)!(l-m)!\right]^{1/2}
\left[2^l\left({l+m\over 2}\right)!\left({l-m\over 2}\right)!\right]^{-1}.
\eqno(2.3)$$
Here the symbol $(-)^k$ is to be interpreted as zero when $k$ is not an
integer. In equation (2.2), the $l=0$ and $l=1$ terms can be be dropped,
since they are not relevant for tidal deformation.
Although our equations are valid for general $l$, throughout this paper
we shall only consider the leading quadrupole term, $l=2$, for which the
relevant coefficients $W_{lm}$ are
$$W_{20}=-\left({\pi\over 5}\right)^{1/2},~~~
W_{2\pm 1}=0,~~~W_{2\pm 2}={1\over 2}\left({6\pi\over
5}\right)^{1/2}.\eqno(2.4)
$$

In the linear approximation, the
effect of the tidal potential on star $M$ is specified by
the Lagrangian displacement $\bxi(\br,t)$ of a fluid element from its
unperturbed position. The equation of motion can be written as
$$\left(\rho{\partial^2\over\partial t^2}+{\cal L}\right)\,\bxi
=-\rho\nabla U,\eqno(2.5)$$
where $\rho$ is the density, and ${\cal L}$ is an operator which specifies
the internal restoring forces of the star.
We now analyze $\bxi(\br,t)$ in to the normal modes of the star
$$\bxi(\br,t)=\sum_\alpha a_\alpha(t)\bxi_\alpha(\br),\eqno(2.6)$$
where $\alpha$ specifies the ``quantum number'' of a normal mode. The
mode eigenfunction $\bxi_\alpha$ satisfies
$$({\cal L}-\rho\omega_\alpha^2)\,\bxi_\alpha(\br)=0,\eqno(2.7)$$
where $\omega_\alpha$ is the angular frequency of the eigenmode.
For a spherical star, the normal modes are labeled by the spherical
harmonic indices, $l,~m$, and by a ``radial quantum number'' $n$,
i.e., $\{\alpha\}=\{n,l,m\}$. For convenience, the indices $\alpha$ and
$\{n,l,m\}$ will be used interchangeably throughout the paper.
For a particular spheroidal mode, the eigenfunction
can be written as the sum of the radial and tangential pieces
$$\bxi_\alpha(\br)=\bxi_{nlm}(\br)
=[\xir_{nl}(r)\be_r+r\xip_{nl}(r)\nabla]Y_{lm}(\theta,\phi),\eqno(2.8)$$
where $\be_r$ is the unit radial vector.
Note that toroidal modes, which are characterized by identically vanishing
Eulerian variation of density (e.g., Cox 1980, p222), are not excited by
potential forces such as the tidal interaction considered here.
Using the orthogonal relations for $\bxi_\alpha$ and the normalization
$$\int\!\!d^3x\rho\,\bxi_\alpha^\ast\cdot\bxi_{\alpha'}=\delta_{\alpha\alpha'},
\eqno(2.9)$$
equation (2.5) can be reduced to a set of equations for $a_\alpha(t)$:
$$\ddot a_\alpha+\omega_\alpha^2 a_\alpha={GM'W_{lm}Q_{nl}\over D^{l+1}}
e^{-im\Phi(t)},\eqno(2.10)$$
where we have defined a {\it tidal coupling coefficient}
$$\eqalign{
Q_{nl} &=\int\!\!d^3x\rho\bxi_{nlm}^\ast\cdot\nabla
\left[r^lY_{lm}(\theta,\phi)\right]\cr
&= \int_0^R\!\rho l r^{l+1}dr\left[\xir_{nl}(r)+(l+1)\xip_{nl}(r)\right],\cr
}\eqno(2.11)$$
(see Zahn 1970, PT). Since $\bxi$ is real, we have $a_{m=2}^\ast=a_{m=-2}$.

Equation (2.10) then essentially determines the response of the star
to the tidal potential, provided we know the functions $D(t)$
and $\Phi(t)$ which specify the decaying orbit.
Using the quadrupole formula for
the gravitational radiation of two point masses, the rate of
change of the orbital separation is given by
$$\dot D=-{64G^3\over 5c^5}{MM'(M+M')\over D^3}.\eqno(2.12)$$
The use of this point-mass formula is adequate, since the effect of tidal
interaction on the orbit is small (see \S 6;
a fully self-consistent formalism, incorporating
the feedback of the tidal interaction on the orbital trajectory,
is discussed in Lai 1994).
It is convenient to define an orbital decay time $t_D$ as
$$t_D\equiv {D\over |\dot D|}={5c^5\over 64G^3}{D^4\over MM'(M+M')}.
\eqno(2.13)$$
The orbital phase function $\Phi(t)$ is
$$\Phi(t)=\int^t\!dt\,\Delta\Omega=\int^t\!dt\,(\Omega-\Omega_s),\eqno(2.14)$$
where $\Omega_s$ is the spin of the star, and $\Omega$ is the orbital angular
frequency, for which we have
$$\Omega^2={G(M+M')\over D^3},~~~\dot\Omega=-{3\dot D\over 2D}\Omega.
\eqno(2.15)$$
When $\Omega_s\neq 0$, the normal modes of the star become much
more complicated, especially when $\Omega_s$ becomes comparable to the
mode frequencies. Throughout our paper, we shall assume
$\Omega>>\Omega_s$, so that the eigenmodes can be adequately approximated
by those of a non-rotating spherical star, for which
$\omega_\al=\omega_{nl}$ does not depend on $m$.
Kochanek (1992), and Bildsten and Cutler (1992) have shown that the
viscosity is too small for significant spin-up of the neutron star
during the inspiral. We point out that even in the
absence of fluid viscosity, angular momentum can be transferred to the star
due to dynamical tidal lag. In \S 6, we will show that such dynamical spin-up
of the neutron star is also negligible.

It is convenient to define a function $b_\alpha(t)$ such that
$$a_\alpha(t)=(GM'W_{lm}Q_{nl})b_\alpha(t)e^{-im\Phi(t)}.\eqno(2.16)$$
Equation (2.10) then becomes
$$\ddot b_\alpha-2im\Omega\dot b_\alpha+
(\omega_\alpha^2-m^2\Omega^2-im\dot\Omega)b_\alpha={1\over D^{l+1}}.
\eqno(2.17)$$
Note that apart from the mode frequency, $b_\al(t)$
does not depends on the properties of the mode and the structure of the star.
The mode properties enter through the tidal coupling coefficient $Q_{nl}$.
Thus the whole problem is divided into two parts: Find $b_\al(t)$ by solving
equation (2.17) for a given $\omega_\alpha$, and calculate the
frequencies and tidal coupling coefficients $Q_{nl}$ for different modes.

The natural units of mass, length, time, and energy in this problem are
$M$, $R$, $(R^3/GM)^{1/2}$, and $GM^2/R$ respectively. We shall use
these units as well as real physical units throughout the paper, whichever
is more convenient. Using the natural units to
nondimensionalize $Q_{nl}$, $\bxi_\alpha$ and $b_\alpha$, we have
$$\bxi(\br,t)=h_o\sum_\alpha W_{lm}Q_{nl}{b_\alpha(t)\over (R/D)^{l+1}}
e^{-im\Phi(t)}\bxi_\alpha(\br),\eqno(2.18)$$
(only $l=2$ terms are included in the sum),
where $h_o$ is simply the typical equilibrium tide height
$$h_o \equiv R\left({M'\over M}\right)\left({R\over D}\right)^3.\eqno(2.19)$$
The equilibrium tide corresponds mainly to the f-mode distortion, for which
$b_\al\sim (R/D)^3$ and $Q_{nl}\sim 1$, as we shall see later.

%%%%%%%%%%%%%%%%%%%%%%%%%%%%%%%%%%%%%%%%%%%%%%%%%%%%%%%%%%%%%%%%%%%%%%
\bigskip
%\vfil\eject
\centerline{\bf 3. DIMENSIONLESS DYNAMICAL TIDAL AMPLITUDE}
\medskip

In this section, we consider the dimensionless tidal amplitude $b_\al(t)$
for a given mode frequency $\omega_\alpha$. This determines the
dynamical aspects of the tidal excitation.

%=======================================================
\bigskip
\centerline{\bf 3.1 Solution Prior to The Resonance}
\medskip

 From equation (2.4), we see that $b(t)$ is
nonzero only for $m=0$ or $m=\pm 2$.
For $m=0$, the tidal potential is quasi-static, and equation (2.17) becomes
$$\ddot b_\alpha+\omega_\alpha^2 b_\alpha={1\over D^{l+1}},
{}~~~~~~~~(m=0).\eqno(3.1)$$
When the orbital decay timescale $t_D$ is much longer than the oscillation
period $1/\omega_\al$, the function $b_\alpha(t)$ is approximately
equal to the quasi-equilibrium value $b_{\alpha,eq}$:
$$b_\al (t)\simeq b_{\alpha,eq}(t)={1\over \omega_\alpha^2D(t)^{l+1}},
{}~~~~~~~~(m=0).\eqno(3.2)$$

Now consider the dynamical tide ($m=\pm 2$). Resonance occurs when
$\omega_\alpha=2\Delta\Omega\simeq 2\Omega$,
at a {\it resonance radius\/} given by
$${D_\alpha\over R}=\left[{4(1+q)GM\over \omega_\alpha^2 R^3}\right]^{1/3},
\eqno(3.3)$$
where $q=M'/M$ is the mass ratio.
Before reaching this resonance, when
$\omega_\alpha^2-(2\Omega)^2>>|\dot\Omega|$, an approximate solution for
$b_\al(t)$ can be found. Dropping the $\ddot b_\alpha$ and $\dot b_\alpha$
terms in equation (2.17), we obtain a zero-th order solution
$$b_\al^{(0)}={1\over D^{l+1}(\omega_\al^2-m^2\Omega^2)}.\eqno(3.4)$$
Thus the zero-th order term in $\dot b_\al$ is
$$\dot b_\al \simeq \left[-(l+1){\dot D\over D}+{2m^2\Omega\dot\Omega\over
\omega_\al^2-m^2\Omega^2}\right]b_\al.\eqno(3.5)$$
Substituting (3.5) into equation (2.17), and dropping the term
$\ddot b\sim b/t_D^2$, we obtain an approximate solution for $b_\al(t)$,
valid for $\omega_\al^2-4\Omega^2>>|\dot\Omega|$:
$$b_\al(t)\simeq {1\over D^{l+1}}\left[\omega_\al^2-m^2\Omega^2
+i\left(2(l+1)m\Omega{\dot D\over D}-m\dot\Omega-{4m^3\Omega^2\dot\Omega\over
\omega_\al^2-m^2\Omega^2}\right)\right]^{-1}.\eqno(3.6)$$

It is important to note that even without viscous
dissipation, the orbital decay
naturally provides a {\it dynamical tidal lag\/}.
This is because the star does not respond
instantaneously to the changing tidal potential.
For $\omega_\al>>\Omega$, this lag angle is
$$\delta_\al \sim {\Delta\Omega\over\omega_\al^2t_D},\eqno(3.7)$$
(recall $\Delta\Omega\simeq \Omega$, since the spin is assumed to be small).
This should be compared with the standard viscosity-induced tidal lag
(cf.~eq.~[1.2])
$$\delta_{visc} \sim {\Delta\Omega\over\omega_\al^2t_{visc}}.\eqno(3.8)$$
For most likely viscosities (see \S 8), we have $t_D<<t_{visc}$,
thus the viscous tidal lag is negligible compared to the dynamical lag induced
by
the changing tidal potential.

%=======================================================
\bigskip
\centerline{\bf 3.2 Resonance Amplitude}
\medskip

As the orbital separation decreases, $b_\al(t)$ increases according
to equation (3.6). We can estimate how close to resonance this equation can
still be valid, from which we
can obtain a scaling relation for the maximum amplitude during a resonance.

Suppose equation (3.6) starts to break down when
$2\Omega \sim (1-\eps)\omega_\al$.
Note that $\eps$ is not determined by $\omega_\al^2-4\Omega^2\sim
\dot\Omega$, which gives $\eps\sim t_{orb}/t_D$ (evaluated at the resonance
radius), where $t_{orb}=2\pi/\Omega$ is the orbital period;
rather, equation (3.6) breaks down at a larger orbital radius, when
the second term (the absolute value) in equation (2.17) becomes comparable to
the third term, or, when the real part and the imaginary part in
equation (3.6) becomes comparable, i.e.,
$${4m^3\Omega^2\dot\Omega\over \omega_\al^2
-m^2\Omega^2}\sim \omega_\al^2-m^2\Omega^2.\eqno(3.9)$$
%where we have assumed $\eps\omega^2>>\dot\Omega$.
Therefore, the critical ``closeness'' to the resonance, before which equation
(3.6) can be applied, is given by
$$\eps\sim \left({\dot\Omega\over \omega_\al^2}\right)_\al^{1/2}
\sim \left({t_{orb}\over t_D}\right)_\al^{1/2},\eqno(3.10)$$
where the subscript ``$\al$'' implies values at the resonance radius $D_\al$.
Evaluating equation (3.6) at $2\Omega\sim (1-\eps)\omega_\al$, we obtain
the amplitude at this near-resonance radius
$$b_\alpha\sim {1\over \eps D_\alpha^{l+1}\omega_\alpha^2}
\sim {1\over\omega_\al D_\al^{l+1}}\left({1\over\dot\Omega}\right)_\al^{1/2}.
\eqno(3.11)$$
The maximum amplitude $|b_{\al,max}|$ during a resonance is of the same order.
 From equation (3.2) for the equilibrium tide, we have
$$\left|{b_{\al,max}\over b_{\al,eq}}\right|_\al
\sim {1\over \eps}\sim \left({t_D\over t_{orb}}\right)_\al^{1/2}.
\eqno(3.12)$$

The maximum amplitude attained during a resonance can also be calculated as
follows. Before the resonance is reached, $a_\alpha(t)$ is essentially
forced to oscillate at frequency $2\Delta\Omega$ (cf.~eqs.~[2.10],
[2.14], [3.6]). But after the resonance,
we expect $a_\alpha(t)$ to oscillate at its own eigenfrequency $\omega_\al$.
Thus, for post-resonance oscillation, instead of equation (2.16), it is
more appropriate to define a function $c_\al(t)$ such such
$$a_\alpha(t)=(GM'W_{lm}Q_{nl})c_\alpha(t)e^{-is_m\omega_\al t},\eqno(3.13)$$
where $s_m=1$ for $m=+2$ and $s_m=-1$ for $m=-2$.
Thus we have
$$b_\al (t)=c_\al (t) e^{im\Phi-is_m\omega_\al t}.\eqno(3.14)$$
Substitution of (3.13) into equation (2.10) yields
$$\ddot c_\al-2is_m\omega_\al \dot c_\al={1\over D^{l+1}}
e^{-im\Phi +is_m\omega_\al t}.\eqno(3.15)$$
Integrating (3.15) with respect to time, and neglecting the $\dot c_\al$ term,
we obtain the post-resonance tidal amplitude
$$c_\al\simeq -{1\over 2is_m\omega_\al}\int\!\!dt {1\over D(t)^{l+1}}
e^{-im\Phi +is_m\omega_\al t}.\eqno(3.16)$$
When $t_D>>1/\omega_\al$, we can approximately set the upper (lower)
limit of the integral to $\infty$ ($-\infty$), and the integral can be
evaluated using the stationary phase approximation.
The maximum amplitude is then given by
$$|c_{\al,max}|\simeq {1\over 2\omega_\al D_\al^{l+1}}
\left({\pi\over\dot\Omega}\right)_\al^{1/2},~~~~~~(m=\pm 2).\eqno(3.17)$$
This has the same scaling as equation (3.11). Using equations (2.12),
(2.15), (3.3) and (3.17), we have
$$\eqalign{
|c_{\al,max}| &\simeq {1\over 16}\left({5\pi\over 6}\right)^{1/2}
\omega_\al^{-5/6}\left({Rc^2\over GM}\right)^{5/4}
q^{-1/2}\left({2\over 1+q}\right)^{5/6}		\cr
&={1\over 2^{21/4}}\left({5\pi\over 6}\right)^{1/2}
\left({D_\al\over R}\right)^{5/4}
\left({Rc^2\over GM}\right)^{5/4}q^{-1/2}\left({2\over 1+q}\right)^{5/4}  \cr
&=5.42 \left(D_{\al,10}R_{10}M_{1.4}^{-1}\right)^{5/4}q^{-1/2}
\left({2\over 1+q}\right)^{5/4},	\cr
}\eqno(3.18)$$
where $D_{\al,10}=D_\al/(10R)$, $R_{10}=R/(10~{\rm km})$, $M_{1.4}
=M/(1.4 M_\odot)$, and $M_\odot$ is the solar mass.
We see that the dimensionless resonance amplitude is larger for
larger values of $D_\al/R$ and $Rc^2/(GM)$. This is because the orbital decay
timescale is larger for larger
$Dc^2/(GM)$ (cf.~eq.[2.13]), and therefore the binary spends a longer time
at a particular resonance radius.
For neutron stars, the ratio $Rc^2/(GM)$ ranges from $4$ to $8$
(e.g., Arnett \& Bowers 1977); for white dwarfs, $Rc^2/(GM)\sim 10^3-10^4$.

%=======================================================
\bigskip
\centerline{\bf 3.3 Numerical Solutions}
\medskip

The dimensionless tidal amplitude $b_\al(t)$ can be determined more accurately
by numerical intergration. Define
$$b_\al(t)=b_\al^{(r)}(t)+ib_\al^{(i)}(t),\eqno(3.19)$$
where $b_\al^{(r)}$ and
$b_\al^{(i)}$ are real. Equation (2.17) then gives
$$\eqalign{
\ddot b_\al^{(r)}+2m\Omega\dot b_\al^{(i)}+(\omega_\al^2-m^2\Omega^2)
b_\al^{(r)}+m\dot\Omega b_\al^{(i)}&={1\over D^{l+1}},\cr
\ddot b_\al^{(i)}-2m\Omega \dot b_\al^{(r)}-m\dot\Omega b_\al^{(r)}+
(\omega_\al^2-m^2\Omega^2)b_\al^{(i)}&=0.\cr
}\eqno(3.20)$$
Numerical integration can be started at a distance larger than the resonance
radius, where $(D-D_\al)/D_\al>>\eps$ (cf.~eq.[3.10]). The initial
condition for $b_\al^{(r)}$ and $b_\al^{(i)}$ can be obtained from
equation (3.6), i.e.,
$$\eqalign{
b_\al^{(r)} &={1\over D^{l+1}(\omega_\al^2-m^2\Omega^2)},\cr
\dot b_\al^{(r)}&=\left[-(l+1){\dot D\over D}+{2m^2\Omega\dot\Omega\over
\omega_\al^2-m^2\Omega^2}\right]b_\al^{(r)}, \cr
b_\al^{(i)}&={1\over \omega_\al^2-m^2\Omega^2}(2m\Omega \dot b_\al^{(r)}
+\dot\Omega b_\al^{(r)}),\cr
\dot b_\al^{(i)}&\simeq 0.\cr
}\eqno(3.21)$$
Note that solutions for $m=2$ and $m=-2$ are related by
$$b_{m=2}^{(r)}=b_{m=-2}^{(r)},~~~~b_{m=2}^{(i)}=-b_{m=-2}^{(i)}.\eqno(3.22)$$
Thus both $m=-2$ and $m=2$ modes are equally excited.

In Figure 1 we show a typical example of the numerical result.
We see that for larger orbital separation, $D>D_\al$, the tidal amplitude
indeed evolves according $a_\al(t)\propto e^{-im\Phi}$
(cf.~eqs.[2.16], [3.6]). Past the resonance, the tidal amplitude approximately
evolves according to $a_\al(t)\propto e^{-is_m\omega_\al t}$.
The absolute amplitude $|c_\al(t)|$ or $|b_\al(t)|$ does not exactly
stay at a constant value after the mode is excited. This is because
the tidal force continues to act upon the oscillation mode,
although the net energy transfer is negligible after the resonance.
Indeed, from our numerical
results, we find that the average post-resonance amplitude $|c_\al(t)|$
agree very well with equation (3.18), except when $D_\al/R \lo 3-4$.

Also note that the maximum amplitude given by equation (3.18)
is not attained instantaneously at $D=D_\al$. From equation (3.16),
the ``duration of the resonance'' $\delta t_\al$, during which the
resonance is effective, is given by
$$\delta t_\al\sim \left|\int\!\!dt e^{-im\Phi +is_m\omega_\al t}\right|
\simeq \left({\pi\over\dot\Omega}\right)_\al^{1/2}
=\left({1\over 3}t_{orb}t_D\right)_\al^{1/2},
\eqno(3.23)$$
(see also RG2).
The corresponding change in the orbital separation $\delta D_\al$ is
$$\eqalign{
\delta D_\al &\sim |\dot D\,\delta t|_\al
\simeq \left({t_{orb}\over 3t_D}\right)_\al^{1/2}D_\al	\cr
&\simeq 0.05 D_\al \left[q\left({1+q\over 2}\right)^{1/2}
\left(D_{\al,10}R_{10}^{-1}M_{1.4}\right)^{5/2}\right]^{1/2},	\cr
}\eqno(3.24)$$
This expression agrees with the numerical result shown in Figure 1.

%%%%%%%%%%%%%%%%%%%%%%%%%%%%%%%%%%%%%%%%%%%%%%%%%%%%%%%%%%%%%%%%%%%%%%
\bigskip
%\vfil\eject
\centerline{\bf 4. EQUATION OF STATE AND CONVECTIVE STABILITY}
\smallskip
\centerline{\bf OF NEUTRON STARS}
\medskip

Having understood the dynamical aspect of the problem
of tidal excitations in \S 3, we now proceed to discuss the properties of
the g-mode oscillations of neutron stars.

The restoring force for g-mode
oscillations in the core of a cold neutron star is the buoyancy due
to the chemical composition
gradient, as first identified by RG1. However, the calculations of
RG1 were based on an approximate (and inconsistent)
ansatz for the Brunt-V\"ais\"al\"a frequency.
In this section, we discuss how g-mode oscillations
depend on the properties of nuclear matter, and how
the Brunt-V\"ais\"al\"a frequency can be obtained self-consistently from
some recent microscopic equations of state (EOS).
Our calculations of g-modes will be presented in \S 5.

%=======================================================
\bigskip
\centerline{\bf 4.1 Composition Gradient and Convective Stability}
\medskip

The existence of stellar g-modes is closely related to the convective stability
(stable stratification)
of the star. For a sufficiently old neutron star, we can assume zero
temperature and zero entropy throughout the star. Thus
buoyancy can only arise from a composition gradient. Consider
a blob of matter in equilibrium with its surrounding, with
pressure $P$, the number of electrons per baryon $Y_e$, and density
$\rho(P,Y_e)$. Imagine the blob
is displaced upwards (against the gravity) adiabatically by a distance $dr$,
where the surrounding matter has pressure $P'$, composition $Y_e'$, and
density $\rho(P',Y_e')$. The blob is in pressure equilibrium with the
surroundings, but its composition is still $Y_e$ during the adiabatic process,
since the timescale for the weak interaction, which is responsible for
the change of $Y_e$, is much longer than the dynamical timescale.
The displaced blob density is therefore $\rho(P',Y_e)$. Convective
stability requires
$$\rho(P',Y_e)>\rho(P',Y_e'),\eqno(4.1)$$
which gives the stability criterion (the Ledoux criterion)
$$\left({\partial\rho\over\partial Y_e}\right)_P
\left({dY_e\over dr}\right)<0.\eqno(4.2)$$

The gravity mode oscillation frequencies are closely related to
the Brunt-V\"ais\"al\"a frequency $N$, defined as
$$N^2=g^2\left({1\over c_e^2}-{1\over c_s^2}\right),\eqno(4.3)$$
(e.g., Cox 1980). Here $g=|{\bf g}|$
is the local gravitational acceleration, $c_s$
is the {\it adiabatic sound speed\/}, as given by
$$c_s^2=\left({\partial P\over\partial\rho}\right)_{Y_e},\eqno(4.4)$$
(the subscript ``s'' means ``adiabatic'', which in this case implies
constant composition), and $c_e$ is given by
$$c_e^2={dP\over d\rho},\eqno(4.5)$$
(the subscript ``e'' stands for ``equilibrium'').
We can easily rewrite the Brunt-V\"ais\"al\"a frequency (eq.~[4.3]) as
$$N^2=g^2\left({\partial\rho\over\partial Y_e}\right)_P
\left({dY_e\over dP}\right)
=-{g\over\rho}\left({\partial\rho\over\partial Y_e}\right)_P
\left({dY_e\over dr}\right),\eqno(4.6)$$
where we have used the hydrostatic equilibrium condition
$dP/dr=-\rho g$. Thus we see that the stability criterion (4.2) is
equivalent to $N^2>0$, or
$$c_s^2-c_e^2=-\left({\partial P\over\partial Y_e}\right)_\rho
\left({dY_e\over d\rho}\right)>0,~~~~~~~({\rm convective~stability}).
\eqno(4.7)$$

We can now easily understand the existence of crustal g-modes (Finn 1987)
and core g-modes (RG1) in a cold neutron star.
In the outer crust, the matter consists
of an electron gas embedded in a lattice of nuclei.
As $\rho$ increases, the nuclei become more neutron-rich
(Baym, Pethick \& Sutherland 1971), so that $dY_e/d\rho<0$\footnote{$~^2$}
{Note that the change of $Y_e$ in the crust is actually discontinuous.}.
On the other hand, since the pressure mainly comes from the electron gas,
for a given total density, the pressure increases as the electron fraction
increases, i.e., $(\partial P/\partial Y_e)_\rho>0$.
Therefore we have $c_s^2-c_e^2>0$, and crustal g-modes are implied.

In the core of a neutron star, the signs of $dY_e/d\rho$ and
$(\partial P/\partial Y_e)_\rho$ are opposite to those in the outer crust.
Consider a simple free npe (neutron, proton and electron) gas
model for the core (e.g., Shapiro \& Teukolsky 1983, RG1).
The electron fraction $Y_e$ increases with density, $dY_e/d\rho>0$. But since
the pressure is mainly provided by the neutrons, we have
$(\partial P/\partial Y_e)_\rho<0$. Again we see $N^2>0$.
More detailed discussions of these points based on general EOS of nuclear
matter are given in \S 4.2.

%=======================================================
\bigskip
\centerline{\bf 4.2 Nuclear Equation of State and Brunt-V\"ais\"al\"a
Frequency}
\medskip

We see from \S 4.1 that calculations of the core g-modes of a neutron star
requires knowledge of both $c_s$ and $c_e$, in addition to the density-pressure
relation. Most of the early EOS's (e.g., Arnett \& Bowers 1977)
do not provide enough information to calculate the sound speed and
to determine the Brunt-V\"ais\"al\"a frequency.
Here we discuss how to obtain these information from a general
parametrization of nuclear matter. Such parametrization
has been used in several recent microscopic nuclear EOS's.

Consider the high density liquid core of a neutron star
consisting of neutrons, protons and electrons
in $\beta-$equilibrium. In general, as a function of the baryon number
density $n$ and the proton fraction $x=n_p/n$, the energy per baryon
of the nuclear matter can be written as
(Lagaris \& Pandharipande 1981; Prakash, Ainsworth \& Lattimer 1988;
Wiringa, Fiks \& Fabrocini 1988, hereafter WFF)
$$E_n(n,x)=T_n(n,x)+V_0(n)+V_2(n)(1-2x)^2,\eqno(4.8)$$
where
$$T_n(n,x)={3\over 5}{\hbar^2\over 2m_n}(3\pi^2n)^{2/3}[x^{5/3}
+(1-x)^{5/3}],\eqno(4.9)$$
is the Fermi kinetic energy of the nucleons, $m_n$ is the nucleon mass.
The last two terms in equation (4.8)
represent contributions due to nucleon interactions: $V_0$ mainly specifies
the bulk compressibility of the matter (hence the pressure-density relation),
and $V_2$ (the ``symmetry potential'') is related to
the symmetry energy of nuclear matter\footnote{$~^3$}
{The symmetry energy $E_s$ is defined such that
$E_n(n,x)=E_n(n,1/2)+E_s(n)(1-2x)^2+\cdots$.
Therefore, $E_s(n)=5T_n(n,1/2)/9+V_2(n)$.
The experimental value is $E_s\sim 30$ MeV at the saturation density $n_s=0.16$
fm$^{-3}$.},
which plays an important role in determining the equilibrium proton fraction.
The parametrization of dense nuclear matter by equation (4.8) is
expected to be generic, and it can simulate many recent
microscopic calculations.

The energy per baryon of relativistic electrons is given by
$$T_e(n,x_e)={3\over 4}\hbar c(3\pi^2n)^{1/3}x_e^{4/3},\eqno(4.10)$$
where $x_e=n_e/n=Y_e$. Charge neutrality requires
$x=x_e$. The total energy per baryon in the liquid core is then
$$E(n,x)=E_n(n,x)+T_e(n,x_e).\eqno(4.11)$$

Minimization of $E(n,x)$ with respect to $x$, i.e., $\partial E/\partial x=0$,
yields the condition for $\beta$-equilibrium
$$\mu_n-\mu_p=\mu_e,\eqno(4.12)$$
where $\mu_i$'s are the chemical potentials of the three species of particles,
with
$$\mu_n-\mu_p=-{\partial E_n\over \partial x}
=4(1-2x)V_2+{\hbar^2\over
2m_n}(3\pi^2n)^{2/3}\left[(1-x)^{2/3}-x^{2/3}\right],\eqno(4.13)$$
and
$$\mu_e=\hbar c(3\pi^2n)^{1/3}x_e^{1/3}.\eqno(4.14)$$
The equilibrium proton fraction $x(n)=x_e(n)$
can then be obtained by solving equations (4.12)-(4.14).
The mass density and pressure are determined via
$$\eqalign{
\rho(n,x)&=n[m_n+E(n,x)/c^2],\cr
P(n,x)&=n^2{\partial E(n,x)\over \partial n}
={2n\over 3}T_n+{n\over 3}T_e+n^2\left[V_0'+V_2'(1-2x)^2\right],\cr
}\eqno(4.15)$$
where $V_0'=dV_0/dn$, etc.
The equilibrium EOS is then obtained when $\rho$ and $P$ are evaluated at
the equilibrium composition $x=x(n)$.
The adiabatic sound speed $c_s$ is given by
$$\eqalign{
c_s^2 &={\partial P\over\partial\rho}
={n\over \rho+P/c^2}{\partial P\over \partial n}	\cr
&={n\over\rho+P/c^2}\left\{{10\over 9}T_n
+{4\over 9}T_e+2n\left[V_0'+V_2'(1-2x)^2
\right]+n^2\left[V_0^{\prime\prime}+V_2^{\prime\prime}(1-2x)^2\right]
\right\}.	\cr
}\eqno(4.16)$$
The difference between $c_s^2$ and $c_e^2$ is given by
$$\eqalign{
c_s^2-c_e^2&={n\over\rho+P/c^2}\left({\partial P\over\partial n}
-{dP\over dn}\right)=-{n\over\rho+P/c^2}
\left({\partial P\over\partial x}\right){dx\over dn}\cr
%&=-{n^3\over\rho}
%\left[\left({\partial^2E_n\over \partial x\partial n}\right){dx\over dn}
%+\left({\partial^2E_e\over \partial x_e\partial n}\right)
%{dx_e\over dn}\right]\cr
&=-{n^3\over\rho+P/c^2}\left[{\partial\over \partial n}(\mu_e+\mu_p-\mu_n)
\right]{dx\over dn}.\cr
}\eqno(4.17)$$
 From equation (4.12), we have
$${dx\over dn}=-\left[{\partial\over \partial n}(\mu_e+\mu_p-\mu_n)
\right]\left[{\partial\over \partial x}(\mu_e+\mu_p-\mu_n)
\right]^{-1}.\eqno(4.18)$$
Equation (4.17) then becomes
$$c_s^2-c_e^2={n^3\over\rho+P/c^2}
\left[{\partial\over \partial n}(\mu_e+\mu_p-\mu_n)
\right]^2\left[{\partial\over \partial x}(\mu_e+\mu_p-\mu_n)
\right]^{-1}.\eqno(4.19)$$
Clearly, convective stability (cf.~eq.[4.7]) requires
$\partial (\mu_e+\mu_p-\mu_n)/\partial x>0$, i.e.,
$${1\over 3x}\mu_e+8V_2+{2\over 3}{\hbar^2\over 2m_n}(3\pi^2n)^{2/3}
\left[(1-x)^{-1/3}+x^{-1/3}\right]>0.\eqno(4.20)$$
Thus unless $V_2$ is extremely negative,
core g-modes always exist. On the other hand, zero-temperature neutron
stars must always be convectively stable, since for the cold nuclear matter
in the ground state, there is no energy available to sustain convective motion
\footnote{$~^4$}{I thank Ed Salpeter and Andreas Reisenegger for pointing this
out to me.}.
Therefore equation (4.20) provides a constraint
on the nuclear equilibrium state.

Finally, we note that just above the nuclear density,
when $\mu_e$ exceeds the muon rest mass $m_\mu$,
muons are energetically allowed via $n\leftrightarrow p+\mu$.
The $\beta$-equilibrium condition becomes
$$\mu_n-\mu_p=\mu_e=\mu_\mu
=[m_\mu^2 c^4+\hbar^2 c^2(3\pi^2nx_\mu)^{2/3}]^{1/2},\eqno(4.21)$$
with $x=x_e+x_\mu$. The expressions derived in this section can
similarly be modified. However, there are greater uncertainties
in our knowledge of nuclear EOS, especially $V_2(n)$.
For definiteness, we will not include muons in our calculations.

%=======================================================
\bigskip
\centerline{\bf 4.3 Neutron Star Core EOS Models}
\medskip

In this paper, four nuclear EOS's will be considered.
The first three are based on recent microscopic calculations of WFF.
For $n\ge 0.07$ fm$^{-3}$, the functions $V_0(n)$ and
$V_2(n)$ obtained from different nucleon Hamiltonians
have been tabulated in Table IV of WFF.
However, to facilitate smooth derivatives of $V_0$ and $V_2$ with
respect to $n$ (which are needed to determine $c_s$
and especially $c_s^2-c_e^2$), we find it more convenient
to use approximate fitting formulae, which we give below:

(1) Model AU: This is the EOS based on nuclear potential AV14+UVII of WFF.
$V_0$ and $V_2$ (in MeV) are fitted as
$$\eqalign{
V_0&=-43+330\,(n-0.34)^2,\cr
V_2&=21\,n^{0.25},\cr
}\eqno(4.22)$$
where $n$ is the baryon number density in fm$^{-3}$.

(2) Model UU: This is based on potential UV14+UVII of WFF.
The fitting formulae are (in the same units as in eq.[4.22])
$$\eqalign{
V_0&=-40+400\,(n-0.3)^2,\cr
V_2&=42\,n^{0.55}.\cr
}\eqno(4.23)$$

(3) Model UT: This is based on potential UV14+TNI of WFF\footnote{$~^5$}
{The results tabulated in Table IV of WFF contain slight errors
(Wiringa 1993, private communication). We have used the
updated results provided by Bob Wiringa.}.
The fitting formulae are (in the same units as in eq.[4.22])
$$\eqalign{
V_0&=-42+350\,(n-0.28)^2,\cr
V_2&=18-130\,(n-0.29)^2.\cr
}\eqno(4.24)$$

Once the functions $V_0(n)$ and $V_2(n)$ are given,
all the relevant EOS parameters needed for the calculations of g-modes
can be obtained using the expressions in \S 4.2. Some of the results are shown
in Figure 2.

The above fitting formulae are valid only for $n\lo 1$ fm$^{-3}$.
This is adequate for our purpose, since we will be mainly
concerned with $M=1.4M_\odot$ NS's, for which all three EOS's yield
a central density that is less than $1$ fm$^{-3}$.
For higher density, both AU and UU
violate causality, indicating considerable
theoretical uncertainties. We note that for AU,
the fitted function $V_2$ does not
reflect the exact behavior of $V_2$ in Table IV of WFF (e.g., $V_2$ in
that table shows non-monotonic behavior for
$0.07~{\rm fm}^{-3}<n<0.7~{\rm fm}^{-3}$, while our fitted
formula is monotonic in $n$). However, there is much greater uncertainty
in the three-body nucleon interaction potential (especially its
isospin dependence) that determines $V_2(n)$ itself.
Also note that for model UT, the quantity $c_s^2-c_s^2$ becomes negative
at very high density. This implies that the model is
not reliable in such a high density regime (see discussion following
eq.~[4.20]).

(4) Model UU2: Our fourth EOS has the same $P(n)$, $\rho(n)$,
and $c_e(n)$ as UU. However, we shall adopt the ansatz
similar to that of RG1 for $c_s^2-c_e^2$ in order to determine to what extent
the g-mode properties are affected.
This ansatz is based on the following consideration.

For free npe gas
(i.e., set $V_0=V_2=0$ for the equations in \S 4.2), $\beta$-equilibrium
gives (see eqs.[4.12]-[4.14])
$$\eqalign{
x &=3\pi^2n\left({\hbar\over 2m_nc}\right)^3[(1-x)^{2/3}-x^{2/3}]^3
\simeq 3\pi^2n\left({\hbar\over 2m_nc}\right)^3		\cr
&\simeq 5.6\times 10^{-3}{\rho\over 2.8\times 10^{14}~{\rm g/cm}^3},\cr
}\eqno(4.25)$$
where the second equality follows from $x<<1$.
The sound speed is given by (eq.[4.16] with $V_0=V_2=0$)
$$c_s^2\simeq {1\over m_n}{\partial P\over \partial n}
\simeq {2\over 3m_n}{\hbar^2\over 2m_n}(3\pi^2n)^{2/3}
\left(1-{x\over 2}\right).\eqno(4.26)$$
 From equations (4.13)-(4.14) and (4.17), we have
$$c_s^2-c_e^2=-{1\over m_n}{dx\over dn}\left[{n\over 3}\mu_e+{2n\over 3}
(\mu_p-\mu_n)\right].
\eqno(4.27)$$
Using equations (4.12), (4.26)-(4.27), we then have
$$c_s^2-c_e^2\simeq {1\over 3m_n}{\hbar^2\over 2m_n}(3\pi^2n)^{2/3}x
\simeq {x\over 2}c_e^2\simeq {x\over 2}c_s^2.\eqno(4.28)$$
This agrees with the expressions given in RG1 (see their eqs.~[28]-[29]).

Now the {\it ansatz\/} for UU2 is to use equation (4.28)
to obtain $c_s$ from $c_e$ and
to use $x(n)$ from equation (4.25),
although both equations are only valid for
free npe gas EOS model\footnote{$~^6$}
{RG1 have also considered the ansatz using more realistic $x(n)$ in equation
(4.28). This is also not a consistent procedure.}.
This ansatz is clearly not a consistent procedure to
obtain $c_s^2-c_e^2$ and hence
the Brunt-V\"ais\"al\"a frequency. In a realistic description of
nuclear matter, the sound speed $c_s$ is determined by the
compressibility of the matter, while the quantity $c_s^2-c_e^2$ is mainly
related to the symmetry energy. Therefore, equation (4.28)
cannot hold in general. Indeed, from Figure 2 we see that
the ratio $(c_s^2-c_e^2)/c_s^2$ in UU2 is qualitatively
different from that in UU, although the other quantities are the same.
Nevertheless, in view of the great uncertainties
in our knowledge of $V_2(n)$, it is worth exploring the g-mode
properties based on this ansatz.

%%%%%%%%%%%%%%%%%%%%%%%%%%%%%%%%%%%%%%%%%%%%%%%%%%%%%%%%%%%%%%%%%%%%%%
\bigskip
%\vfil\eject
\centerline{\bf 5. CORE g-MODE OSCILLATIONS AND TIDAL}
\smallskip
\centerline{\bf COUPLING COEFFICIENTS}
\medskip

We now discuss our calculations of the g-mode oscillations of a isolated
cold neutron star
based on the EOS models of \S 4. The resulting mode frequencies $\omega_\al$
and the tidal coupling coefficients $Q_{nl}$ (eq.~[2.11]) will be
used in \S 6 to determine the energy transfer and angular momentum transfer
due to resonant excitations in coalescing neutron star binaries.

%=======================================================
\bigskip
\centerline{\bf 5.1 Basic Equations}
\medskip

In this paper, we use Newtonian linearized
equations to calculate the oscillation modes.
The use of Newtonian equations is consistent with our Newtonian description of
tidal interactions (\S 2). For f-mode, general relativistic effects
are expected to modify our results of oscillation frequencies by not more than
$GM/(Rc^2)\sim 20\%$. Linear oscillations of nonrotating stars in
Newtonian theory have been discussed extensively in the literature
(e.g., Ledoux 1974, Cox 1980). Here, for completeness and to clarify the
notations, we give some of the relevant  equations below.

The basic equations that govern the oscillations of stars are the
equations of hydrodynamics,
i.e., the continuity equation, the equation of motion, energy equation
(which specifies the adiabatic nature of the oscillations), and the Poisson
equation for the gravitational potential.
The linearized equations can be written in terms of the radial
component of the fluid displacement $\xi^r$ (cf.~eq.[2.8]),
the Eulerian perturbations of density, pressure and gravitational potential,
$\delta\rho$, $\delta P$ and $\delta\phi$.
Separating out the angular dependence $Y_{lm}(\theta,\phi)$,
these linearized equations are (here and throughout this section, the
mode index $\al=\{n,l,m\}$ is suppressed)
$$\eqalign{
{\partial \over\partial r}(r^2\xi^r) &={\rho g\over \Gamma_1 P}(r^2\xi^r)
+\left[{l(l+1)\over \omega^2}-{r^2\rho\over\Gamma_1 P}\right]
\left({\delta P\over\rho}\right)+{l(l+1)\over\omega^2}\delta\phi,\cr
{\partial \over\partial r}\left({\delta P\over\rho}\right) &=
{\omega^2+Ag\over r^2}(r^2\xi^r)-A\left({\delta P\over\rho}\right)
-{\partial \over\partial r}(\delta\phi), \cr
{1\over r^2}{\partial \over\partial r}\left(r^2{\partial\over \partial r}\delta
\phi\right) &={l(l+1)\over r^2}\delta\phi+4\pi G\delta\rho,\cr
}\eqno(5.1)$$
where $\omega$ is the eigenfrequency of the mode, and
$$\Gamma_1={\rho c_s^2\over P}, ~~~~A=-{N^2\over g},~~~~
\delta\rho ={\rho^2\over\Gamma_1 P}\left({\delta P\over\rho}\right)
-{\rho A\over r^2}(r^2\xi^r).
\eqno(5.2)$$
Here, $c_s^2$, $N^2$ can be calculated using equations (4.3), (4.16)-(4.17).
The transverse component of the displacement vector
(see eq.~[2.8]) is given by
$$\xi^\perp(r)={1\over\omega^2 r}\left[{\delta P(r)\over\rho}+\delta\phi(r)
\right].
\eqno(5.3)$$

The eigenmodes can be determined by solving equation (5.1) with
appropriate boundary conditions.
For $r\rightarrow 0$, regularity requires
$$r^2\xi^r={l\over\omega^2}(Y_o+\Phi_o)r^{l+1},
{}~~~{\delta P\over\rho}=Y_or^l,~~~
\delta\phi=\Phi_o r^l,~~~~~~~~(r\rightarrow 0),
\eqno(5.4)$$
where $Y_o$ and $\Phi_o$ are constants.
At the stellar surface, the Lagrangian perturbation of the pressure,
$\Delta P$, must vanish, i.e.,
$$\Delta P=\delta P-\rho g\xi_r=0,~~~~~~(r=R).\eqno(5.5)$$
Also, the gravitational potential and force must be continuous
across the surface. This gives
$${\partial \over\partial r}\delta\phi+{l+1\over r}\delta\phi
=-4\pi G\rho\xi_r,~~~~~~(r=R).
\eqno(5.6)$$

Numerically, we fix $Y_o=1$, which simply sets a scale for the
eigenfunctions. Equations (5.1) are integrated outward from $r\rightarrow 0$
with boundary conditions (5.4), using a fifth-order Runge-Kutta scheme.
The outer boundary conditions at $r=R$ (eqs.[5.5]-[5.6]) can be matched
using a standard shooting routine (Press et al. 1992), therefore
the eigenvalues $\omega^2$ and $\Phi_o$ can be obtained after an initial guess.
The number of nodes in the functions is counted to determine the orders
of the modes. After $\omega^2$ is obtained for a specific mode,
the eigenfunctions are normalized according to equation (2.9), i.e.,
$$\int_0^R\!\rho r^2dr\left[\xi^r(r)^2+l(l+1)\xi^\perp (r)^2\right]=1.
\eqno(5.7)$$
Then the tidal coupling coefficient $Q_{nl}$ is calculated using
equation (2.11).
To test our numerical code, we have done calculations for
polytropic stellar models, with $P=K\rho^\Gamma$, where the polytropic
index $\Gamma$ and the adiabatic index $\Gamma_1>\Gamma$ are
constant throughout the star. For a few specific values of $\Gamma$ and
$\Gamma_1$, the mode frequencies and
$Q_{nl}$ have already been tabulated in the literature
(e.g., Lee \& Ostriker 1986). Our results for the polytropic models agree
with these tabulated values.

We construct equilibrium neutron star models by solving
Newtonian hydrostatic equations. We use Newtonian equations
for consistency with our use of Newtonian equations in calculating
oscillation modes and in describing the tidal interaction.
The EOS's we have used
are those discussed in \S 4.2-4.3 for $n\ge 0.07$ fm$^{-3}$,
where the pressure,
sound speed, and $c_s^2-c_e^2$ (hence the Brunt-V\"ais\"al\"a frequency)
are all calculated self-consistently (except model UU2).
For lower density crustal regions, we use the Baym, Bethe \& Pethick (1971) EOS
above neutron drip ($\rho\go 4\times 10^{11}$ g/cm$^3$), and then join onto the
Baym, Pethick \& Sutherland (1971) EOS below neutron drip. In all of our
calculations, we set $c_s^2=c_e^2$ in the crustal region, therefore
effectively suppressing the crustal g-modes while concentrating on the core
g-modes. Since the density in the crustal region is much smaller,
from equation (2.11) we see that the tidal coupling coefficients for these
crustal modes are likely to be very small (as found by Shibata 1993, who
used highly simplified neutron star models).

%=======================================================
\bigskip
\centerline{\bf 5.2 Numerical Results and Discussions}
\medskip

Table 1 gives our numerical results for the oscillation
frequencies and tidal coupling coefficients $Q_{nl}$ of the f-mode and
the first two g-modes for neutron stars based on the four EOS models
discussed in \S 4.3. Note that since the gravitational wave frequency
is twice that of the orbital frequency $\Omega$,
and resonances occur when $\omega_\al=2\Omega$,
the listed mode frequencies also correspond to the gravitational wave
frequencies at resonances.
We see that the four neutron star models considered
have very similar stellar radii for the given canonical mass $M=1.4M_\odot$,
and their f-mode properties are also very similar.
This is because the four EOS's of \S 4.3
predict very similar bulk properties (e.g., pressure, compressibility)
for the nuclear matter. However, the g-mode properties are very different
for the four models. In particular, although UU and UU2 yield the
identical neutron star mass-radius relations,
the g-mode frequencies predicted by them
differ by large factors. These differences reflect the sensitive dependence of
g-modes on the symmetry energy of the nuclear matter.

 From Table 1 we see that the frequencies of g-modes
are much smaller than that of
f-mode. This is a direct consequence of the small difference between
$c_s$ and $c_e$ (cf.~Fig.2). Also, $|Q_{nl}|$ becomes smaller
as the radial order of modes increases. This is because of the strong
cancellation in the integral for $Q_{nl}$ (eq.~[2.11]), as already
noted by Cowling (1941). Indeed, using the equation of continuity,
$$\delta\rho_\al=-\nabla\cdot(\rho\bxi_\al),\eqno(5.8)$$
the expression for $Q_{nl}$  (eq.~[2.11]) can be rewritten as
$$Q_{nl}=\int_0^R\!r^{l+2}\delta\rho_{nl}(r) dr.
\eqno(5.9)$$
Since the order of a mode corresponds to the number of nodes in
$\delta\rho_{nl}(r)$, we clearly see that for the higher-order mode
$|Q_{nl}|$ is smaller. Based on our numerical values of $Q_{nl}$ for the
first few g-modes, we find that for all NS models considered, $|Q_{nl}|$
decreases as $\omega_{nl}$ decreases faster than
$|Q_{nl}|\propto \omega_{nl}^{3.3}$.

The frequencies of the first ten quadrupole g-modes are given in Figure 3.
They generally lie in the range of $10~$Hz to $100~$Hz.
An approximate asymptotic expression for the frequencies
can be easily obtained.
For high-order g-modes ($n>>1$), the perturbation of the
gravitational potential
$\delta\phi$ can be neglected (``Cowling approximation''). Eliminating $\delta
P/\rho$ from the first two equations in (5.1), and noticing that
$\omega\rightarrow 0$, we have
$${\partial^2\over\partial r^2}\left(r^2\xi^r\right)
\simeq {l(l+1)\over\omega^2}{\partial\over\partial r}\left(
{\delta P\over\rho}\right)\simeq
-{l(l+1)\over\omega^2r^2}N^2\left(r^2\xi^r\right).\eqno(5.10)$$
Assuming $r^2\xi^r\sim e^{ikr}$, we obtain the WKB local dispersion
relation for high-order g-modes
$$k(r)^2\simeq {l(l+1)\over\omega^2 r^2}N(r)^2.\eqno(5.11)$$
The eigenvalues for $\omega^2$ are then determined by the stationarity
condition
$$\int_0^R\!k(r)dr=(n+C)\pi,\eqno(5.12)$$
where $C$ is a constant of order unity. Thus, asymptotically,
the high-order g-mode angular frequencies are given by
$$\omega_{nl}\simeq {\sqrt{l(l+1)}\over (n+C)\pi}
\int_0^R\!{N(r)\over r}dr.\eqno(5.13)$$
%For more detailed discussion of the asymptotic behaviors of non-radial
%stellar oscillations, see, e.g., Tassoul (1980).

The values of $Q_{nl}$ for the higher-order g-modes are harder to calculate
numerically as they become much smaller. Indeed, in evaluating the values of
$Q_{nl}$, it is necessary to know the eigenfunctions of the g-modes
very precisely, since a small f-mode contamination can significantly
change the obtained values (Reisenegger 1994).
To obtain meaningful numbers of $Q_{nl}$ for the g-modes,
it is also important to implement self-consistent
hydrodynamical equations in the numerical calculations,
so that the Hermiticity of the oscillation
operator ${\cal L}$ is maintained. For example, we have found that
significant systematic errors in $Q_{nl}$ can be introduced if
$c_e^2=dP/d\rho$ is not satisfied to a high accuracy when interpolating
tabulated EOS. For the g-modes given in Table 1, we have checked that
the orthogonality relation (2.9) is satisfied to sufficient accuracy,
therefore our values of $Q_{nl}$ for these g-modes are not contaminated by
the f-mode.

We have also done calculations of the oscillation modes
using the Cowling approximation. We find, as expected, that
the g-mode frequencies are nearly unaffected. However, for a given stellar
model, the Cowling approximation leads to a larger f-mode frequency
(by $\sim 50\%$) and larger values of $|Q_{nl}|$ for the first few g-modes
(by as large as $80\%$).

%%%%%%%%%%%%%%%%%%%%%%%%%%%%%%%%%%%%%%%%%%%%%%%%%%%%%%%%%%%%%%%%%%%%%
\bigskip
%\vfil\eject
\centerline{\bf 6. TIDAL ENERGY AND ANGULAR MOMENTUM}
\medskip

In this section, we combine our results of \S 3 and \S 5 to calculate the
amplitude of distortion, energy transfer and angular momentum transfer to the
neutron star due to tidal interaction.
The effects on the orbital phase will be considered in \S 7.

%=======================================================
\bigskip
\centerline{\bf 6.1 Tidal Distortion Amplitude}
\medskip

A measure of the tidal distortion of the neutron
star is the quantity $a_\al$ (cf.~eq.[2.6]).
For equilibrium, quasi-static tide ($m=0$), equation (3.2) gives
$$|a_{\al,eq}|\simeq q\,\omega_{nl}^{-2}|W_{20}Q_{nl}|
\left({R\over D}\right)^3.	\eqno(6.1)$$
where $\omega_{nl}$ is the mode angular frequency in units of
$(GM/R^3)^{1/2}$.
The dominant tidal distortion comes from the f-mode, for which
$\omega_{nl}\sim 1$ and $Q_{nl}\sim 1$ (see Table 1), thus
$$|a_{0,eq}|\sim {h_0\over R}= q\,\left({R\over D}\right)^3,
\eqno(6.2)$$
where $h_0$ is the typical height of the tidal bulge
(see eq.~[2.19]).
% While g-mode oscillations mainly involve horizontal
% motion of the fluid elements, it is mainly the f-mode oscillation
% that produce deformation of the star from the spherical shape.

Now consider the dynamical tides ($m=\pm 2$).
Before the resonance, the distortion amplitude associated with
a particular mode increases as the orbit decays (see eq.~[3.6]).
After the resonance, the amplitude stays
nearly constant. From equations (2.16) and (3.18), the tidal distortion
amplitude resulting from the resonant excitation is given by
$$\eqalign{
|a_{\al,max}| &\simeq {\pi\over 32}
\omega_{nl}^{-5/6} |Q_{nl}| \left({Rc^2\over GM}\right)^{5/4}q^{1/2}
\left({2\over 1+q}\right)^{5/6}	\cr
&=2.7\times 10^{-3}\left({f_{nl}\over 100 {\rm Hz}}\right)^{-5/6}
\left({|Q_{nl}|\over 0.0003}\right)M_{1.4}^{-5/6}q^{1/2}
\left({2\over 1+q}\right)^{5/6},\cr
}\eqno(6.3)$$
where $f_{nl}=\omega_{nl}/2\pi$ is the mode
frequency, which is also equal to the frequency of the
gravitational wave when the resonance occurs.
The reference values $f_{nl}=100$ Hz and $|Q_{nl}|=0.0003$ have been chosen
to approximately agree with the lowest order g-mode in model UU (see Table 1).
Note that equation (6.3) is valid for both $m=2$ and $m=-2$,
and it is half of the result given by RG2 (see their eq.~[14])
\footnote{$~^7$}
{In the PT formalism used by RG2, the contributions from $m=2$ and $m=-2$
terms are not equal, while it is clear from our analysis (see eqs.~[2.10],
[3.16] or [3.22]) that the $m=2$ and $m=-2$ modes contribute to tidal
distortion and energy equally. The difference
comes from the fact that contributions
from $m=2$ and $m=-2$ modes have been re-grouped to derive the final
expressions of PT. Thus, although explicit evaluations in RG2 are made
only for $l=m=2$, RG2's results actually include contributions from
both the $m=2$ and $m=-2$ modes, since the other term in PT's
formula are much smaller.}.
Since $|a_\al|<<1$, the linear approximation we have adopted throughout the
paper is valid.

%=======================================================
\bigskip
\centerline{\bf 6.2 Tidal Energy Transfer}
\medskip

Next we look at tidal energy transferred to the neutron
star. The kinetic energy of the star is given by
$$E_k(t) ={1\over 2}\int\!\!d^3x\,\rho
{\partial\bxi^\ast\over\partial t}\!\cdot\!{\partial\bxi\over\partial t}
={1\over 2}\sum_\al |\dot a_\al|^2,
\eqno(6.4)$$
where we have used equations (2.6) and (2.9).
Similarly, using equation (2.7),
the potential energy (including the internal energy) of the oscillations
is given by
$$E_p(t) ={1\over 2}\int\!\!d^3x\,\bxi^\ast\!\cdot\!{\cal L}\!\cdot\,\bxi
={1\over 2}\sum_\al\omega_\al^2|a_\al|^2,
\eqno(6.5)$$
(e.g., Shapiro \& Teukolsky 1983, p141-143).
The total tidal energy of the neutron star is
$E=E_k+E_p$. Substituting equations (2.16) and (3.13), and
using the natural units introduced at the end of \S 2, we have
$$\eqalign{
E_k(t) &={1\over 2}\sum_\al (M'W_{lm}Q_{nl})^2 |\dot b_\al-im\Omega b_\al|^2
={1\over 2}\sum_\al (M'W_{lm}Q_{nl})^2 |\dot c_\al-is_m\omega_\al c_\al|^2, \cr
E_p(t) &={1\over 2}\sum_\al (M'W_{lm}Q_{nl})^2 \omega_\al^2|b_\al(t)|^2
={1\over 2}\sum_\al (M'W_{lm}Q_{nl})^2 \omega_\al^2|c_\al(t)|^2.
}\eqno(6.6)$$

Another way to consider the tidal energy of the neutron star is
to look at the rate of energy transfer:
$$\dot E=-
\int\!d^3x\,\rho\,{\partial \bxi\over\partial t}\cdot\nabla U^\ast.
\eqno(6.7)$$
Using equations (2.2), (2.6), (2.11) and (2.16), we have
$$\dot E=\sum_\al (M'W_{lm}Q_{nl}){1\over D^{l+1}}
e^{im\Phi}\dot a_\al
=\sum_\al (M'W_{lm}Q_{nl})^2{1\over D^{l+1}}
(\dot b_\al-im\Omega b_\al).
\eqno(6.8)$$
It is easy to check that $\dot E$ in this expression is real
(as it should be), since
the contributions to the imaginary part from the $m=+2$ term and that from
the $m=-2$ term cancel.
Moreover, using equation (2.10), we can show that
this expression for $\dot E$ is consistent with
equations (6.4)-(6.5) for $E_k$ and $E_p$.

For equilibrium, quasi-static tide ($m=0$),
when orbital decay time $t_D>>1/\omega_{nl}$,
the kinetic tidal energy is small compared to the potential
energy (thus ``quasi-static''). The equilibrium tidal energy associated with a
particular mode is then
$$E_{\al,eq}={1\over 2}{GM^2\over R}q^2\omega_{nl}^{-2}(W_{20}Q_{nl})^2
\left({R\over D}\right)^6.
\eqno(6.9)$$
Again, the dominant equilibrium tidal energy comes from the f-mode
$$E_{0,eq}\sim {GM^2\over R}q^2\left({R\over D}\right)^6,
\eqno(6.10)$$
which is the expected result (e.g., Lai et al 1993).

Now consider the energy transfer $\Delta E_{nl}$ due to the resonant
excitation of a specific g-mode.
Since $c_\al(t)$ is approximately constant after the resonance, the
kinetic energy and the potential energy are equal.
Equations (3.18) and (6.6) then yield
$$\eqalign{
\Delta E_{nl} &\simeq 2\omega_{nl}^2 |a_{\al,max}|^2	\cr
&\simeq {\pi^2\over 512}\left({GM^2\over R}\right)
\omega_{nl}^{1/3} Q_{nl}^2\left({Rc^2\over GM}\right)^{5/2}
q\left({2\over 1+q}\right)^{5/3}.\cr
}\eqno(6.11)$$
Note that the factor of $2$ in the first equality comes about because
the $m=2$ and $m=-2$ terms contribute equally to the excitation energy, and
we have included both of them in (6.11).
The ratio of $\Delta E_{nl}$ to the orbital energy
$$E_{orb}\simeq -{GMM'\over 2D},\eqno(6.12)$$
is given by
$$\eqalign{
{{\Delta E_{nl}}\over |E_{orb}|}
&\simeq {\pi^2\over 128}\omega_{nl}^{-1/3}Q_{nl}^2
\left({Rc^2\over GM}\right)^{5/2}q^2\left({2\over 1+q}\right)^{4/3} 	\cr
&= 1.0\times 10^{-6}\left({f_{nl}\over 100~{\rm Hz}}\right)^{-1/3}
\left({|Q_{nl}|\over 0.0003}\right)^2 M_{1.4}^{-7/3}R_{10}^2\,
q^2\left({2\over 1+q}\right)^{4/3},	\cr
}\eqno(6.13)$$
where $|E_{orb}|$ is evaluated at $D=D_{nl}$.
This ratio plays an important role in determining the effect
of resonant energy transfer on the orbital decay rate (see \S 7).
Since $\Delta E_{nl}<<|E_{orb}|$,
the resonance has only a small effect on the orbit, and the use of the
trajectory of ideal two point masses (cf.~eq.[2.12])
is a good approximation.
Our result for $\Delta E_{nl}$ agree with
that of RG2 (see their eqs.~[15] and [17]), but note that our equations
(6.11) and (6.13) include both $m=2$ and $m=-2$ modes (see footnote 7).

%=======================================================
\bigskip
\centerline{\bf 6.3 Angular Momentum Transfer and Dynamical Spin-up of the
Neutron Star}
\medskip

Associated with the tidal energy transfer is an angular momentum transfer
to the neutron star. This is possible because of the dynamical tidal lag
induced by the decaying orbit, as already noted in \S 3.1
(see eqs.~[3.6]-[3.7]).
The $z-$component of the tidal torque on the neutron star
per unit mass is given by
$$\tau_z=-{\bf e}_z\cdot (\br\times\nabla)U=-{\partial \over\partial \phi}U.
\eqno(6.14)$$
Thus the total torque is
$$\cN_z =\int\!d^3x (\rho+\delta\rho)\tau_z^\ast
=\int\!d^3x\delta\rho\tau_z^\ast.
\eqno(6.15)$$
Using the continuity equation (5.8), we have
$$\eqalign{
\cN_z &=-\sum_\al a_\al(t) \int\!d^3x \nabla\cdot(\rho\bxi_\al)
\,\tau_z^\ast		\cr
&=\sum_\al (M'W_{lm}Q_{nl})^2{1\over D^{l+1}}(-imb_\al)
=\sum_\al (M'W_{lm}Q_{nl})^2{1\over D^{l+1}}mb_\al^{(i)}.\cr
}\eqno(6.16)$$
Note that this is non-zero since the contribution from the $m=+2$ term and
that from the $m=-2$ term add up (cf.~eq.[3.22]).
Clearly, the equilibrium tide ($m=0$) does not contribute to
angular momentum transfer.

Prior to the resonance, the tidal torque can be evaluated using equation
(3.6). In order of magnitude, the torque associated with a specific mode is
$$\cN_{z,\al}\sim {GM^2\over R}q^2\omega_{nl}^{-2}Q_{nl}^2
\left({R\over D}\right)^6\delta_\al,
\eqno(6.17)$$
where the dynamical tidal lag angle $\delta_\al$ is given by equation (3.7).
The dominant f-mode contribution is
$$\cN_{z,0}\sim {GM^2\over R}q^2 \left({R\over D}\right)^6 \delta_0
\sim {GM^2\over R}q^2 \left({R\over D}\right)^6 {\Omega\over\omega_0^2 t_D}.
\eqno(6.18)$$
It is easy to see that this non-resonant
dynamical tidal torque does not appreciably
affect the neutron star spin. In particular, this torque is
not sufficient to synchronize the spin with the orbital angular velocity:
the timescale for synchronization is given by
$$t_{syn}\sim {MR^2\Omega\over \cN_{z,0}}
\sim q^{-2}\left({D\over R}\right)^6t_D,
\eqno(6.19)$$
which is much larger than the orbital decay time $t_D$ unless $D$ becomes
comparable to $R$.

Now consider the total angular momentum transferred to the star,
$\Delta J_\al$, due to a particular resonance.
 From equation (6.16), this is given by
$$\Delta J_\al=(M'W_{lm}Q_{nl})^2\int\!dt{1\over D^{l+1}}(-imb_\al),
\eqno(6.20)$$
where only the real part should be included (recall that
the imaginary part from the $m=2$ term and that from the $m=-2$ term cancel).
Using equations (3.14)-(3.15), we have
$$\eqalign{
\Delta J_\al &=(M'W_{lm}Q_{nl})^2 (-im)\int\!dt{1\over D^{l+1}}c_\al
e^{im\Phi-i\omega_\al s_m t}	\cr
&=(M'W_{lm}Q_{nl})^2 (-im)\int\!dt\,c_\al
(\ddot c_\al^\ast+2i\omega_\al s_m \dot c_\al^\ast).\cr
}\eqno(6.21)$$
Integrate by parts, and notice that
the term proportional to $c_\al\dot c_\al^\ast$ can be neglected,
since after the resonance, $c_\al$ is nearly constant.
Adding up contributions from both $m=2$ and $m=-2$ terms,
the total angular momentum transfer $\Delta J_{nl}$ due to the resonance
is given by
$$\Delta J_{nl}\simeq 2(M'W_{lm}Q_{nl})^2\omega_{nl} |m||c_{\al,max}|^2
=4\omega_{nl} |a_{\al,max}|^2.
\eqno(6.22)$$
Using equation (6.3), we have
$$\Delta J_{nl}
\simeq {\pi^2\over 256}(GM^3R)^{1/2}\omega_{nl}^{-2/3}Q_{nl}^2
\left({Rc^2\over GM}\right)^{5/2}q\left({2\over 1+q}\right)^{5/3}.
\eqno(6.23)$$
Note that the energy transfer and angular momentum transfer are related by
$$\Delta J_{nl} ={\Delta E_{nl}\over \Omega_{nl}}
={2\Delta E_{nl}\over\omega_{nl}},
\eqno(6.24)$$
where $\Omega_{nl}=\omega_{nl}/2$ is the orbital angular velocity at
the resonance.

The effect of this resonant angular momentum transfer on the spin of the
neutron star can be considered if we
{\it assume\/} that the angular momentum goes into uniform rotation of the
star. The resulting change of the spin rate
due to the resonance is then given by
$$\Delta\Omega_{s,{nl}}={\Delta J_{nl}\over I},\eqno(6.25)$$
where $I=2\kappa MR^2/5$ is the moment of inertia of the star, $\kappa<1$
is of order unity. The ratio of $\Delta\Omega_{s,{nl}}$ and $\Omega_{nl}$ is
$$\eqalign{
{\Delta\Omega_{s,{nl}}\over \Omega_{nl}}
&={5\pi^2\over 256\kappa}\omega_{nl}^{-5/3}Q_{nl}^2
\left({Rc^2\over GM}\right)^{5/2}q\left({2\over 1+q}\right)^{5/3}	\cr
&=1.5\times 10^{-4}\kappa^{-1}\left({f_{nl}\over 100~{\rm Hz}}\right)^{-5/3}
\left({|Q_{nl}|\over 0.0003}\right)^2 M_{1.4}^{-5/3}
\,q\left({2\over 1+q}\right)^{5/3}.	\cr
}\eqno(6.26)$$
Therefore, spin-up due to resonant angular momentum transfer is negligible.
%In the absence of fluid viscosity, the angular momentum transferred
%does not manifest as uniform spin of the star, since the fluid circulation is
%conserved.
In reality, the angular momentum transferred
does not manifest itself as uniform spin of the star, since
the viscous timescale to achieve
uniform rotation is long.
Thus the actual change of spin rate
is even smaller than that given by the above expression.

 %%%%%%%%%%%%%%%%%%%%%%%%%%%%%%%%%%%%%%%%%%%%%%%%%%%%%%%%%%%%%%%%%%%%%
\bigskip
%\vfil\eject
\centerline{\bf 7. EFFECTS ON ORBITAL DECAY RATE: PHASE ERROR}
\medskip

% As discussed in \S 6.3 (see eq.~[6.13]),
% the effect of resonant transfer of energy and angular momentum
% from the binary orbit to the neutron star have small effect
% on the orbital decay rate. However, this small effect may
% nevertheless be important for the gravitational wave signal detection. As
% mentioned in our introduction, small orbital phase error in the theoretical
% wave template can destroy a possible detection using
% the matched filter technique (e.g., Cutler et al 1993).

We now calculate the orbital phase error due to the resonant
energy transfer and angular momentum transfer
from the binary orbit to the neutron star.
This is important because,
as mentioned in the introduction, a small phase error in the theoretical
gravitational wave template can destroy a possible detection using
the matched filter technique.
% (e.g., Cutler et al 1993).

Because of the relation (6.24), a circular orbit will remain circular after
the resonance. Therefore it is sufficient to consider the energy of
the system only. Excluding a constant,
the total energy of the system, $E_{tot}$, can be written as a sum of the
orbital energy $E_{orb}$ and the stellar energy $E$, i.e.,
$E_{tot}=E_{orb}+E$. Thus
$$\dot E_{tot}={dE_{tot}\over dD}\dot D
=\left({d E_{orb}\over dD}+{dE\over dD}\right)\dot D.\eqno(7.1)$$
The number of orbital cycles, $N_{orb}$, can be obtained from
$$dN_{orb}={\Omega\over 2\pi}{dD\over \dot D}
={\Omega\over 2\pi}dD\left({dE_{orb}\over dD}+{dE\over dD}\right)/\dot E_{tot}.
\eqno(7.2)$$
The first term in (7.2) is simply the point mass result.
The second term represents
the error induced by the change of stellar energy:
$$d(\Delta N_{orb})={\Omega\over 2\pi}dD\left({dE\over dD}\right)
/\dot E_{tot}.\eqno(7.3)$$
This is a general expression, which applies to dynamical tidal effects
as well as equilibrium tidal effects.
Now consider the total accumulated error on the orbital cycles
$\Delta N_{orb,nl}$ due to
resonant excitation of a particular mode with angular frequency
$\omega_{nl}$. Since $dE/dD$ is large only
near the resonance radius $D_{nl}$ (see eq.[3.3]), we can integrate equation
(7.3) to obtain
$$\Delta N_{orb,nl}
\simeq {\Omega_{nl}\over 2\pi}{\Delta E_{nl}\over \dot E_{tot}}
=-{t_D\over t_{orb}}{\Delta E_{nl}\over |E_{orb}|},
\eqno(7.4)$$
where $t_D\equiv |E_{orb}/\dot E_{tot}|=|D/\dot D|$ (see eq.~[2.13]),
and $t_{orb}=2\pi/\Omega$ is the orbital period. Note that all the
quantities in equation (7.4)
are evaluated at the resonance radius $D_{nl}$. Here the negative sign
implies that the effect of energy transfer is to make the binary coalesce
slightly faster (i.e., slightly larger $|\dot D|$) compared
with the ideal point mass case.
Substituting equations (2.13) and (6.13) into (7.4), we have
$$\eqalign{
\Delta N_{orb,nl} &\simeq -{5\pi\over 4096}\omega_{nl}^{-2}Q_{nl}^2
\left({Rc^2\over GM}\right)^5 q\left({2\over 1+q}\right) 	\cr
&\simeq -4.3\times 10^{-4} \left({f_{nl}\over 100~{\rm Hz}}\right)^{-2}
\left({|Q_{nl}|\over 0.0003}\right)^2 M_{1.4}^{-4}\,R_{10}^2
\,q\left({2\over 1+q}\right).	\cr
}\eqno(7.5)$$
%The error in the number of cycles of gravitational wave is simply
%$\Delta N_{gw,nl}=2\Delta N_{orb,nl}$.
Note that this phase error does not accumulate instantaneously at $D=D_\al$
(see \S 3.3). From equation (3.23), the number of orbital cycles
that unfold during the period of effective resonance is given by
$$\delta N_{orb,nl}\sim {\Omega_\al\over 2\pi}\delta t_\al
\simeq 12.\left({f_{nl}\over 100~{\rm Hz}}\right)^{-5/6}
M_{1.4}^{-5/6}\,q^{-1/2}\left({2\over 1+q}\right)^{1/6}.
\eqno(7.6)$$

The small size of $\Delta N_{orb,nl}$ in equation (7.5)
implies that {\it the phase errors due to resonant excitations of g-modes
are negligible for constructing gravitational wave templates\/}.
Therefore, to a very good approximation,
binary neutron stars can be
treated as point masses during the inspiraling phase.
Hydrodynamical tidal effects are dominated by the f-mode distortion,
and they are important only when the neutron stars are close to
contact (see eqs.~[6.2], [6.10]).

%Of course, the exact values of $\Delta N_{orb,nl}$
%depend sensitively on $f_{nl}$ and $Q_{nl}$ of the g-modes,
%which in turn depend on the EOS of the neutron star.
%In addition to the self-consistent NS models described in \S 4.3,
%we have also explored g-mode properties based on some other NS models.

 %%%%%%%%%%%%%%%%%%%%%%%%%%%%%%%%%%%%%%%%%%%%%%%%%%%%%%%%%%%%%%%%%%%%%
\bigskip
%\vfil\eject
\centerline{\bf 8. TIDAL HEATING OF BINARY NEUTRON STARS}
\medskip

Having determined in \S 6 the tidal energy in the decaying
binary neutron stars, a natural question concerns where this
energy goes. Without damping mechanisms, this energy is stored in the form
of potential energy and kinetic energy of various
oscillation modes. There are basically three processes that can lead to
damping of g-modes (RG1): gravitational radiation, relaxation toward
chemical equilibrium by neutrino emission, and viscous dissipation.
The damping timescales for all of these mechanisms are likely to be
much larger than the orbital decay time  of interest here, i.e., the last few
minutes prior to the binary merger. However, viscous dissipation leads to
heating of the neutron stars, thus changing the physical properties
of the stars even before merging takes place. Therefore it is of interest
of consider the heating processes in some detail.

The standard estimation of viscous tidal dissipation goes as follows.
The rate of energy dissipation is
$$\dot E_{visc}\sim \nu M \left|{\partial v_i\over\partial x_k}\right|^2,
\eqno(8.1)$$
where $\nu$ is the kinematic shear viscosity (in units of cm$^2/$s),
and $\partial v_i/\partial x_k$ is the strain tensor.
The typical tidal bulge height is $h_0$,
and the bulge rises and falls at angular velocity $2\Delta\Omega$ (see
eq.~[1.2]). If the tidal distortion of the star varies over a length-scale
comparable to the stellar radius, then the shear strain tensor is of order
$\partial v_i/\partial x_k
\sim \Delta\Omega h_0/R$. Thus the viscous dissipation rate is
$$\dot E_{visc} \sim \nu M\left(\Delta\Omega {h_0\over R}\right)^2
={M(h_0\Delta\Omega)^2\over R^2/\nu}
=\nu M (\Delta\Omega)^2q^2\left({R\over D}\right)^6.
\eqno(8.2)$$
This is simply the {\it tidal kinetic energy\/}
$E_k\sim M (h_0\Delta\Omega)^2$ divided by
the viscous time $t_{visc}\sim R^2/\nu$.
For a synchronized system ($\Delta\Omega=0$), there is no viscous dissipation.

It is clear from the above derivation that this simple
relation is mainly for the f-mode tide, i.e., higher-order modes and resonances
have been neglected.
However, higher-order mode oscillations involve more shear
motion of the fluid, thus the length-scale and timescale
for viscous dissipation are smaller for these modes. More importantly,
when g-mode resonances occur,
the viscous dissipation can start at a larger separation,
at which the orbit decay is still slow. As a result, more heat can be
generated during the whole inspiral. By contrast,
dissipation due to the non-resonant
f-mode oscillation is mainly effective for small orbital separation (see
eq.~[8.2]), at which binary decay is fast.

Of course, the amount of viscous dissipation ultimately depends on
the viscosity of the fluid, which is not at all well understood.
Our main intention in this section is to consider the effect of g-mode
resonances on the viscous dissipation rate.
Although the stellar spin can be incorporated in our formalism without much
difficulty, we shall focus on the non-spinning case,
so that $\Delta\Omega=\Omega$. In this case, equation (8.2) indicates
that for non-resonant viscous dissipation $\dot E_{visc}\propto 1/D^9$.

%=======================================================
\bigskip
\centerline{\bf 8.1 Viscous Damping Rate of Normal Oscillation Modes}
\medskip

Here we consider the viscous damping rates of individual modes.
To leading order, viscous stress does not change the oscillation
frequency, but it induces mode damping. This damping effect is equivalent to
adding an imaginary part to the oscillation frequency, i.e.,
$\omega_\al'=\omega_\al+i\gamma_\al.$
The mode energy damping rate is given by
$$2\gamma_\al={\dot E_{visc,\al}\over E_\al}.\eqno(8.3)$$
Here $E_\al$ is the energy of the eigenmode:
$$E_\al=2E_{\al,k}=\omega_\al^2
\int\!\!d^3x\rho\,\bxi_\al^\ast\cdot\bxi_\al=\omega_\al^2.
\eqno(8.4)$$
where we have used the normalization (2.9).
The viscous dissipation rate is given by (suppressing the mode index $\al$)
$$\dot E_{visc}=-\int\!d^3x \sigma_{ik}v_{i,k}^\ast
=-\int\!d^3x \left[{1\over 2}\eta \,|v_{i,k}+v_{k,i}-{2\over 3}\delta_{ik}
\nabla\cdot {\bf v}|^2
+\zeta |\nabla\cdot {\bf v}|^2\right],
\eqno(8.5)$$
(Landau \& Lifshitz 1987)
where $\sigma_{ik}$ is the viscous stress tensor
$$\sigma_{ik}=\eta \left(v_{i,k}+v_{k,i}-{2\over 3}\delta_{ik}\nabla\cdot
{\bf v}\right)+\zeta \delta_{ik}\nabla\cdot {\bf v},\eqno(8.6)$$
$\eta=\rho\nu$ is the dynamical shear viscosity,
and $\zeta$ is the bulk viscosity.

Substituting ${\bf v}=-i\omega\bxi$ and equation (2.8) into equation (8.5),
the damping rate can be evaluated. In the absence of bulk viscosity,
the result has been obtained
by Higgins \& Kopal (1968, see also Kopal 1978).
For $m=0$ modes,
$$\gamma^{(shear)}_{m=0}= {1\over 2}\int_0^R r^2dr \eta F(r),\eqno(8.7)$$
where
$$\eqalign{
F(r)= &2\left({\partial \xi^r\over\partial r}\right)^2
+\left({2\xi^r\over r}-l(l+1){\xi^\perp\over r}\right)^2	\cr
&+l(l+1)\left({\partial\xi^\perp\over\partial r}+{\xi^r-\xi^\perp\over
r}\right)^2
+(l-1)l(l+1)(l+2)\left({\xi^\perp\over r}\right)^2-{2\over 3}G(r)^2,\cr
}\eqno(8.8)$$
and
$$G(r)={\partial\xi^r\over\partial r}+{2\over r}\xi^r-l(l+1){\xi^\perp\over r}.
\eqno(8.9)$$
For modes with $m\neq 0$\footnote{$~^8$}
{Our definition of $Y_{lm}$ is that of Jackson (1975),
which is different from that used in Higgins \& Kopal (1968).
Therefore the results are different by a numerical factor.
We have checked that for the f-mode of incompressible sphere (the
Kelvin mode), our result using equations (8.7)-(8.9)
agrees with the standard one, as first derived by H. Lamb;
see Cutler \& Lindblom (1987).},
$$\gamma^{(shear)}_{m\neq 0}={(l+|m|)!\over (l-|m|)!}\gamma^{(shear)}_{m=0}.
\eqno(8.10)$$
When bulk viscosity is present, we can add
the bulk viscous dissipation to the damping rate, i.e.,
$$\gamma=\gamma^{(shear)}+\gamma^{(bulk)},\eqno(8.11)$$
where
$$\gamma_{m=0}^{(bulk)} ={1\over 2}\int_0^R r^2dr\zeta G(r)^2,~~~~~
\gamma_{m\neq 0}^{(bulk)}={(l+|m|)!\over (l-|m|)!}\gamma_{m=0}^{(bulk)}.
\eqno(8.12)$$

To specify the dependence of $\gamma$ on the mode structure, independent of
the actual values of viscosities, let us consider
viscosities of the form
$$\eta=\eta_N\left({\rho\over\rho_N}\right)^2,~~~~
\zeta=\zeta_N\left({\rho\over\rho_N}\right)^2.\eqno(8.13)$$
where $\rho_N$ is some fiducial density.
Such a density dependence is appropriate for the microscopic viscosity
in the core of neutron star (\S 8.3).
The damping coefficient can be written as (for the $m=0$ mode)
$$\gamma^{(shear)}={\eta_N R\over M}\left({M\over R^3\rho_N}\right)^2
\Sigma^{(shear)},~~~~~
\gamma^{(bulk)}={\zeta_N R\over M}\left({M\over R^3\rho_N}\right)^2
\Sigma^{(bulk)},\eqno(8.14)$$
where $\Sigma$'s are dimensionless coefficients depending only on the mode
eigenfunctions:
$$\Sigma^{(shear)}= {1\over 2}\int_0^R r^2dr \rho^2 F(r),~~~~
\Sigma^{(bulk)}={1\over 2}\int_0^R r^2dr\rho^2 G(r)^2.\eqno(8.15)$$
Here all quantities in the integrands are in units such that $M=R=1$.

The values of $\Sigma^{(shear)}$ and $\Sigma^{(bulk)}$ are given in Table 1
for various modes. Clearly, $\Sigma^{(shear)}$ increases as the order of the
mode increases. For higher-order g-modes, our numerical results
yield an approximate scaling relation,
$\Sigma_{nl}^{(shear)}\propto \omega_{nl}^{-2}$.
%This is expected: the wavefunctions of higher-order modes
%have more nodes, and the length scales for the oscillations are thus smaller.

%=======================================================
\bigskip
\centerline{\bf 8.2 Viscous Dissipation in Coalescing Binaries}
\medskip

With the results of \S 8.1, we can now write down the expressions
for the viscous energy dissipation rate in the coalescing binary neutron
stars. Using equations (2.6), (2.16) and (8.3)-(8.5), we have
$$\dot E_{visc}=
\sum_\al (M'W_{lm}Q_{nl})^2 |\dot b_\al-im\Omega b_\al|^2 2\gamma_\al
=\sum_\al (M'W_{lm}Q_{nl})^2 |\dot c_\al-is_m\omega_\al c_\al|^2
2\gamma_\al.
\eqno(8.16)$$
It is clear that the contribution from each mode in this expression
is simply twice of the tidal kinetic energy (see eq.~[6.6])
of the mode divided by energy damping time of that mode, $1/(2\gamma_\al)$.

Consider first the contribution from the f-mode. In this case, $b_\al$ is given
by equation (3.6). Since $|\dot b_\al|<<\Omega |b_\al|$, we have, after adding
contributions from both $m=2$ and $m=-2$ terms,
$$\dot E_{visc,0}\simeq
2 (M'W_{l\pm 2}Q_{0})^2 |2\Omega b_0|^2 2\gamma_0
\simeq {24\pi\over 5}\left({GM^2\over R}\right)\omega_0^{-4}Q_0^2
\left({R\over D}\right)^92\gamma_0,
\eqno(8.17)$$
where we have assumed $\omega_0>>\Omega$ and have used equation (2.4).
Here and below, $\gamma$ refers to $m=\pm 2$ modes, thus we have, e.g.,
$\gamma^{(shear)}=4!\gamma_{m=0}^{(shear)}=
4!(\eta_NR/M)(M/\rho_NR^3)^2\Sigma^{(shear)}$
(see eqs.~[8.10], [8.14]). Clearly, equation (8.17) agrees with our simple
dimensional analysis in (8.2).

Now consider the dissipation associated with a specific g-mode.
Prior to the resonance,
one can still use equation (3.6) in (8.16) to obtain an expression similar
to (8.17). After the resonance, the tidal energy stored in the mode stays
nearly constant. The viscous dissipation rate is then given by
$$\eqalign{
\dot E_{visc,\al}\simeq &(\Delta E_\al)2\gamma_\al 	\cr
\simeq &3.2\times 10^{-8}\left({GM^2\over R}\right)
\left({f_{nl}\over 100~{\rm Hz}}\right)^{1/3}\left({Q_{nl}\over 0.0003}
\right)^2
R_{10}^3 M_{1.4}^{-8/3}q\left({2\over 1+q}\right)^{5/3} 2\gamma_\al, \cr
}\eqno(8.18)$$
where we have used equation (6.11) for
$\Delta E_\al=\Delta E_{nl}$, and
we have included both $m=2$ and $m=-2$ contributions.

The relative importance of the non-resonant viscous dissipation (8.17)
and the resonant dissipation (8.18) can be compared by calculating
the total viscous dissipation accumulated during the inspiral from
$D=\infty$ to the orbital radius before merging $D=D_m$:
$$E_{visc}=\int\!dt\dot E_{visc}=-\int_\infty^{D_m}
{dD\over D}t_D\dot E_{visc},
\eqno(8.19)$$
where $t_D$ is given by equation (2.13).
For the f-mode non-resonant dissipation, we have
$$\eqalign{
E_{visc,0} &\simeq {3\pi\over 80}\left({GM^2\over R}2\gamma_0{GM\over c^3}
\right)\left({R\over D_m}\right)^5\left({Rc^2\over GM}\right)^4
q{Q_0^2\over \omega_0^4}	\cr
&\simeq 0.037\,q R_{10}^4M_{1.4}^{-4}\left({GM^2\over R}2\gamma_0{GM\over c^3}
\right),
}\eqno(8.20)$$
where in the second equality we have assumed $D_m/R=3$ and have used
$\omega_0^2\simeq 1.5,~Q_0\simeq 0.56$ (see Table 1).
For the g-mode, we can neglect its contribution prior to the resonance, i.e.,
we assume $\dot E_{visc,\al}\simeq 0$ for $D>D_\al$. The total accumulated
viscous dissipation due to the resonant g-mode oscillation is then
$$\eqalign{
E_{visc,\al} &\simeq {5\pi^2\over 128^2}\omega_\al^{-7/3}
Q_{nl}^2\left({Rc^2\over GM}\right)^{13/2}
\left({2\over 1+q}\right)^{4/3}
\left({GM^2\over R}2\gamma_\al {GM\over c^3}\right)	\cr
&\simeq 0.010\,\left({f_{nl}\over 100~{\rm Hz}}\right)^{-7/3}
\left({Q_{nl}\over 0.0003}\right)^2R_{10}^3M_{1.4}^{-16/3}
\left({2\over 1+q}\right)^{4/3}\left({GM^2\over R}2\gamma_\al {GM\over c^3}
\right). \cr
}\eqno(8.21)$$
Because $|Q_{nl}|$ decreases rapidly as $\omega_{nl}$ decreases,
only the resonance with the lowest order g-mode is important.
%Since $\gamma_\al^{(shear)}\propto \Sigma_\al^{(shear)}$,
%and $\Sigma_\al^{(shear)}>\Sigma_0^{(shear)}$, we see that
%$E_{visc,\al}>>E_{visc,0}$.
In obtaining equations (8.20) and (8.21), we have assumed that
$\gamma_\al$ remains constant during the inspiral. In reality, the viscosities
depend on temperature, which increases as the neutron star is heated up.
Nevertheless, since $\gamma_\al^{(shear)}>\gamma_0^{(shear)}$,
equations (8.20) and (8.21) already indicated that
g-mode resonances can be important in heating the neutron star
compared to the standard f-mode non-resonant heating.

%=======================================================
\bigskip
\centerline{\bf 8.3 Neutron Star Viscosity}
\medskip

The expressions derived in \S 8.1 and \S 8.2 are quite general, and can be
used with different viscosities.
To actually determine the amount of heating of the coalescing
binary neutron stars before merging, we need to know the viscosity of the
nuclear matter. Calculations by Flowers \& Itoh (1976, 1979) yield an
estimate for the microscopic shear viscosity in liquid core of neutron star.
Approximate fitting formulae have been given by Cutler and Lindblom (1987).
For normal liquid core, the viscosity is dominated by neutron-neutron
scattering, with
$$\eta\simeq 1.1\times 10^{19}T_8^{-2}\left({\rho\over\rho_N}\right)^{9/4}
{}~{\rm g\,cm}^{-1}{\rm s}^{-1},\eqno(8.22)$$
where $\rho_N=2.8\times 10^{14}$ g/cm$^3$, and $T_8=T/(10^8~{\rm K})$.
When the temperature falls below a critical temperature
$\sim 10^9$ K, both neutron and proton are likely
undergo a phase transition to become a
superfluid (see, e.g., Wambach, Ainsworth \& Pines 1991 for review).
In this low temperature regime, the viscosity is dominated
by electron-electron scattering, and it is given by
$$\eta \simeq  4.7\times 10^{19}T_8^{-2}\left({\rho\over\rho_N}\right)^2
{}~{\rm g\,cm}^{-1}{\rm s}^{-1}.\eqno(8.23)$$
Since coalescing binary neutron stars are extremely old,
their internal heat must have radiated away.
All cooling calculations indicate that neutron stars older than
$10^8$ years have core temperature
less than $\sim 10^6$ K (and surface temperature less than
$10^5$ K) (e.g., Tsuruta 1992).
Therefore, without tidal heating, the cores of the neutron stars in the
coalescing binary systems are likely to be in the superfluid state.

The bulk viscosity of neutron star arises from the delay in achieving
beta equilibrium as the density is changed. Calculation of Sawyer (1989)
yields
$$\zeta\simeq 4.8\times 10^{18}T_8^6\left({\rho\over\rho_N}\right)^2
\left({\omega\over 1~{\rm s}^{-1}}\right)^{-2}
{}~{\rm g\,cm}^{-1}{\rm s}^{-1},\eqno(8.24)$$
where $\omega$ is the angular frequency of the perturbation\footnote{$~^9$}
{This result is based on modified Urca process. When the proton fraction is
sufficiently large to allow for direct Urca process (Lattimer et al.
1991), the value of $\zeta$ is larger than equation (8.24) by a factor of
$(\mu_n/kT)^2\sim 5\times 10^7T_8^{-2}$ (where $\mu_n$ is the Fermi energy
of the neutron). Also, equation (8.24) is valid only when the
the deviation from chemical equilibrium
$\delta\mu=\mu_n-\mu_p-\mu_e$ is much smaller than $kT$. More general
expression can be found in Haensel (1992).}.
Because of the strong temperature dependence, bulk viscosity is important
only when $T>>10^8$ K. We shall neglect it in our following discussions.

Of course, one should be cautious in using these microscopic viscosities.
The stable stratification indicates that turbulent viscosity can not be
present in a cold neutron star.
% (however, note the curves for
%EOS3 in Figure 2, where $c_s^2-c_e^2$ becomes negative at very high density).
Also, tidal distortion is not likely to induce turbulence
and convection in the binary stars, at least in the linear regime
(Seguin 1976). However, it is not completely clear whether any
other ``anomalous'' viscosities, such as the mutual friction between
the electrons and the superfluid neutron vortices
(e.g., Sauls 1989, Mendell 1991), can be present in a cold neutron star.

%=======================================================
\bigskip
\centerline{\bf 8.4 Heating of the Neutron Stars}
\medskip

With the microscopic viscosity given by equation (8.23), the mode damping
constant is (for $m=\pm 2$):
$$\gamma_\al=4\times 10^{-5}T_8^{-2}\Sigma_\al M_{1.4}R_{10}^{-5}
{}~~{\rm s}^{-1}, \eqno(8.25)$$
where we have suppressed the superscript ``shear''.

Below we shall focus on neutron stars with $M=1.4M_\odot$ and $R=10$ km.
Equations (8.17) and (8.25) then give the non-resonant heating rate
$$\dot E_{visc,0}\simeq 2.6\times 10^{50}\left({R\over D}\right)^9T_8^{-2}
{}~~{\rm erg\,s}^{-1}, \eqno(8.26)$$
where we have used $\Sigma_0\simeq 3$ (see Table 1).
Similarly, equations (8.18) and (8.25) yield the resonant heating rate
$$\dot E_{visc,\al}\simeq 1.3\times 10^{42}\left({f_{nl}\over 100~{\rm Hz}}
\right)^{1/3}\left({Q_{nl}\over 0.005}\right)^2 T_8^{-2}\Sigma_\al
{}~{\rm erg\,s}^{-1}. \eqno(8.27)$$

The heat content of the neutron star is mainly given by the
thermal energy of the nonrelativistic degenerate free neutron gas in the core:
$$U\simeq {\pi^2\over 4}Nk_BT{k_BT\over \mu_n} \simeq
4.5\times 10^{45}T_8^2~~{\rm erg}, \eqno(8.28)$$
where $N$ is the total number of nucleons and $k_B$ is the Boltzmann constant
\footnote{$~^{10}$}
{When both neutron and proton become superfluid, their heat capacity
is greatly reduced by a factor of order $e^{\Delta/kT}$, where $\Delta\sim
0.1-1$ MeV is the superfluid gap energy. In this regime, the main
thermal content of the neutron star is due to that of the relativistic
degenerate electrons, and $U$ is smaller by a factor $2x$. Using
equation (4.25) as a simple estimate of $x$, we have
$U\simeq 1.2\times 10^{44}T_8^2~({\rm erg})$. If we use this expression
instead of eq.~(8.28), the resulting temperature is higher than eqs.
(8.30)-(8.31) by a factor of $2.4$.}.
The thermal evolution equation of the neutron star during the orbital decay
can be simply written as
$${dU\over dt}=\dot E_{visc}+\dot E_{cool}.  \eqno(8.29)$$
The cooling term $\dot E_{cool}$ due to neutrino emission and
surface photon emission can be shown to be small at this low temperature and
will be neglected.
If we include only the f-mode non-resonant heating $\dot E_{visc,0}$, and use
equation (8.28), then integrating equation (8.29) yields:
$$T_8 \simeq 0.36\left({3R\over D}\right)^{5/4}.
\eqno(8.30)$$
On the other hand, if we include only the resonant dissipation,
we obtain, for $D<<D_{nl}$,
$$T_8 \simeq 0.42 \left({f_{nl}\over 100~{\rm Hz}}\right)^{-7/12}
\left({|Q_{nl}|\over 0.0003}\right)^{1/2}
\left({\Sigma_{nl}\over 20}\right)^{1/4}.
\eqno(8.31)$$
Thus the resonant excitation of the g-mode can be as important as
the standard f-mode tidal dissipation
in heating the neutron star. Moreover, this resonant excitation
leads to significant heating at larger orbital separation, where
the f-mode tidal heating is small.
Our value of the temperature is much smaller than that estimated by
M\'esz\'aros \& Rees (1992). To obtain their value of $\sim 10^{11}$ K,
we would require a viscosity that is larger than the microscopic
value by a factor of order $10^{10}-10^{12}$, corresponding to a
kinematic viscosity comparable to the maximum possible value in a neutron star
$\nu_{max}\sim (GMR)^{1/2}\sim 10^{16}$ cm$^2/$s.
We consider this rather unlikely.
We therefore expect that before the binary neutron stars merge,
they remain relatively cold and degenerate. Tidal dissipation heats up the
neutron stars from less than $10^6$ K to $\sim 10^8$ K before the stars
come into contact.

%%%%%%%%%%%%%%%%%%%%%%%%%%%%%%%%%%%%%%%%%%%%%%%%%%%%%%%%%%%%%%%%%%%%%
\bigskip
%\vfil\eject
\centerline{\bf 9. CONCLUSIONS}
\medskip

We have studied in this paper various consequences of tidal excitations of
oscillations in coalescing binary neutron stars.
The energy transfer and angular momentum transfer from the orbit to the star
due to the resonant excitations of low-frequency g-modes
are found to be small because of the weak coupling
between the g-mode and the tidal potential. The induced orbital
phase errors are negligible for the detection of
gravitational wave from such binary systems by LIGO and VIRGO.
Therefore, only the f-mode, quasi-equilibrium tide needs to be considered.
Binary models based on ellipsoidal figures
(e.g., Lai, Rasio \& Shapiro 1994a,~1994b)
are thus expected to give a good description of the effects of
tidal interaction on the orbits of coalescing neutron star binaries.
However, we find that the resonant excitation of the lowest-order g-mode
can play a significant role in the tidal heating of
the neutron stars prior to merger.

Our calculations of the core g-mode oscillations of neutron stars
indicate that the g-mode frequencies depend not only on the
pressure-density relation, but also on the symmetry properties of the
nuclear matter. Since the resonant amplitude of the tidal
distortion associated with the lowest-order g-mode can
be as large as $\sim 0.5\%$,
if some electromagnetic signature of this induced
oscillation can be detected, as suggested by RG2, then we may be able to
constrain the properties of nuclear matter in the neutron star.
We caution that the actual values of the g-mode frequencies
obtained in this paper are by no means conclusive.
In our calculations, we have made several
assumptions, e.g., we have used Newtonian theory,
we have neglected the crustal density discontinuities and shear modulus,
the thermal effects, as well as possible superfluid effects.
Also, our calculations are based on a single type of
nuclear equations of state (WFF), since these are the only EOS available to us
from which we can extract enough information to calculate the g-modes
self-consistently (see \S 4). It is desirable that future microscopic
calculations of high density nuclear matter not only provide
the pressure-density relation, but also provide sufficient details
which would allow one to obtain the sound speed $c_s$ and the small
difference $c_s^2-c_e^2$, so that g-mode properties based on a wider variety
of EOS can be calculated.

Finally, we note that although the focus of this paper has been
neutron star binaries, the expressions
we have established in \S 2-3 and \S 6-8 are generally
valid and can be
applied to some other binary systems as well, as long as the tidal
distortion remains small and the linear approximation is valid.
For example, coalescing white dwarf binaries have been considered
as important sources of low-frequency gravitational
waves that are potentially detectable by future space-based interferometers
(Evans, Iben \& Smarr 1987). The restoring force for the g-modes of white
dwarfs is provided by the thermal energy of degenerate electrons
(Baglin \& Heyvaerts 1969, Chanmugam 1972).
Since the orbital decay rate is much smaller for a white dwarf binary,
the effects of g-mode resonances are much more pronounced.
Indeed, using the typical value of the ratio $Rc^2/(GM)\sim 10^3-10^4$
for white dwarfs, and using similar dimensionless values for
$\omega_{nl}$ and $Q_{nl}$ (e.g., $\omega_{nl}\sim 0.1$,
$Q_{nl}\sim 10^{-4}$),
we find that the energy transferred to the white dwarf during a
resonance can be significant compared to the orbital energy
(see eq.~[6.13]). Thus g-mode resonances are important for the orbital
evolution
of coalescing binary white dwarfs.

%%%%%%%%%%%%%%%%%%%%%%%%%%%%%%%%%%%%%%%%%%%%%%%%%%%%%%%%%%%%%%%%%%%%%
\bigskip
\centerline{\bf ACKNOWLEDGEMENTS}
\medskip

It is a pleasure to thank Curt Cutler, Sam Finn,
Edwin Salpeter, Stuart Shapiro, Saul Teukolsky, and especially
Phil Nicholson and Ira Wasserman for useful discussions
on the subjects related to this paper. I also thank
Andreas Reisenegger, Stuart Shapiro and Saul Teukolsky
for reading an early version of this paper and helpful suggestions.
I am particularly grateful to
Andreas Reisenegger for sending his paper to me prior to publication and
making many critical comments on the earlier version of this paper.
Communication with Bob Wiringa concerning nuclear equation of
state is also greatly appreciated, as well as Greg Cook for sharing
computer data files of some equation of states for neutron stars.
This work has been supported in part
by NSF Grant AST 91-19475 and NASA Grant NAGW-2364 to Cornell University.

%%%%%%%%%%%%%%%%%%%%%%%%%%%%%%%%%%%%%%%%%%%%%%%%%%%%%%%%%%%%%%%%%%%%%
%\vfill\eject
\bigskip
\bigskip
\centerline{\bf REFERENCES}
\medskip
\def\bysame{\hbox to 50pt{\leaders\hrule height 2.4pt depth -2pt\hfill .\ }}

\hi{
Abramovici, A., et al. 1992, Science, 256, 325}

\hi{
Arnett, W.D., \& Bowers, R.L. 1977, ApJS, 33, 415}

\hi{
Baglin, A., \& Heyvaerts, J. 1969, Nature, 222, 1258}

\hi{
Baym, G., Bethe, H.A., \& Pethick, C.J. 1971, Nucl. Phys. A175, 225}

\hi{
Baym, G., Pethick, C.J., \& Sutherland, P. 1971, ApJ, 170, 299}

\hi{
Bildsten, L., \& Cutler, C. 1992, ApJ, 400, 175}

\hi{
Bradaschia, C., et al. 1990, Nucl. Instrum. \& Methods, A289, 518}

\hi{
Chanmugam, G. 1972, Nature, 236, 83}

\hi{
Cowling, T.G. 1941, MNRAS, 101, 367}

\hi{
Cox, J.P. 1980, Theory of Stellar Pulsation (Princeton University Press)}

\hi{
Cutler, C., \& Lindblom, L. 1987, ApJ, 314, 234}

\hi{
Cutler, C. et al. 1993, Phys. Rev. Lett., 70, 2984}

%\hi{
%Eichler, D., Livio, M., Piran, T., \& Schramm, D.N. 1988, Nature, 340, 126}

\hi{
Evans, C.~R., Iben, I., \& Smarr, L. 1987, ApJ, 323, 129}

\hi{
Finn, L.S. 1987, MNRAS, 227, 265}

\hi{
Flowers, E., \& Itoh, N. 1976, ApJ, 206, 218}

\hi{
\bysame 1979, ApJ, 230, 847}

\hi{
Goldreich, P., \& Nicholson, P. 1989, 342, 1079}

\hi{
Haensel, P. 1992, A\&A, 262, 131}

\hi{
Higgins, T.P., \& Kopal, Z. 1968, Astro. Space Science, 2, 352}

\hi{
Jackson, J.D. 1975, Classical Electrodynamics, 2nd Ed.
(John Wiley \& Sons: New York), p.~99.}

\hi{
Kochanek, C. S. 1992, ApJ, 398, 234}

\hi{
Kopal, Z. 1978, Dynamics of Close Binary Systems
(D.~Reidel Publishing Company: Dordrecht-Holland), p.~126.}

\hi{
Lagaris, I.E., \& Pandharipande 1981, Nucl. Phys., A369, 470}

\hi{
Lai, D. 1994, Ph.D. thesis, Cornell University}

\hi{
Lai, D., Rasio, F.A., \& Shapiro, S.L. 1993, ApJ, 406, L63}

\hi{
\bysame 1994a, ApJ, 420, 811}

\hi{
\bysame 1994b, ApJ, submitted}

\hi{
Landau, L.D., \& Lifshitz, E.M. 1987, Fluid Mechanics, 2nd
Ed. (Oxford: Pergamon Press)}

\hi{
Lattimer, J.M., Pethick, C.J., Prakash, M., \& Haensel, P. 1991,
Phys. Rev. Lett., 66, 2701}

\hi{
Ledoux, P. 1974, in Stellar Instability and Evolution
(IAU Symposium No.~59), ed. P. Ledoux, A. Noels and A.W. Rodgers
(D.~Reidel Publishing Company: Dordrecht-Holland)}

\hi{
Lee, H.M., \& Ostriker, J.P. 1986, ApJ, 310, 176}

\hi{
McDermott, P.N., Van Horn, H.M., \& Scholl, J.F. 1983, ApJ, 268, 837}

\hi{
McDermott, P.N., Van Horn, H.M., \& Hansen, C.J. 1988, ApJ, 325, 725}

\hi{
Mendell, G. 1991, ApJ, 380, 530}

\hi{
M\'esz\'aros, P., \& Rees, M.~J. 1992, ApJ, 397, 570}

\hi{
Narayan, R., Piran, T., \& Shemi, A. 1991, ApJ, 379, L17}

\hi{
Nicholson, P. 1978, Ph.D. thesis, Caltech}

%\hi{
%Paczy\'nski, B. 1986, ApJ, 308, L43}

%\hi{
%Paczy\'nski, B. 1991, Acta Astron. 41, 257}

\hi{
Phinney, E. S. 1991, ApJ, 380, L17}

\hi{
Prakash, M., Ainsworth, T.L., \& Lattimer, J.M. 1988, Phys. Rev. Lett., 61,
2518}

\hi{
Press, W.~H., Teukolsky, S.~A., Vetterling, W.~T, \&
Flannery, B.~P. 1992,
Numerical Recipes: The Art of Scientific Computing, 2nd Ed.
(Cambridge: Cambridge Univ.\ Press)}

\hi{
Press, W.H., \& Teukolsky, S.A. 1977, ApJ, 213, 183}

\hi{
Reisenegger, A. 1994, Preprint}

\hi{
Reisenegger, A., \& Goldreich, P. 1992, ApJ, 395, 240 (RG1)}

\hi{
\bysame 1994, ApJ, in press (RG2)}

\hi{
Sauls, J.~A. 1988, in Timing Neutron Stars, eds. H. \"Ogelman \&
E.P.J. van den Heuvel (Kluwer: Dordrecht), 457}

\hi{
Sawyer, R.~F. 1989, Phys. Rev. D, 39, 3804}

\hi{
Schutz, B.~F. 1986, Nature, 323, 310}

\hi{
Seguin, F.~H. 1976, ApJ, 207, 848}

\hi{
Shapiro, S. L., \& Teukolsky, S. A. 1983, Black Holes,
White Dwarfs, and Neutron Stars (New York: Wiley)}

\hi{
Shibata, M. 1993, Osaka University preprint}

%\hi{
%Tassoul, M. 1980, ApJS, 43, 469}

\hi{
Thorne, K.~S. 1987, in 300 Years of Gravitation,
eds. S.~W. Hawking \& W. Israel (Cambridge: Cambridge University Press), 330}

\hi{
Tsuruta, S. 1992, in Physics of Isolated Pulsars, eds.
K. Van Riper, R. Epstein \& C. Ho (Cambridge University Press)}

\hi{
Wambach, J., Ainsworth, T.~L., \& Pines 1991, in Neutron Stars: Theory and
Observations, eds. J. Ventura \& D. Pines (Kluwer: Dordrecht)}

\hi{
Wiringa, R. B., Fiks, V., \& Fabrocini, A. 1988, Phys. Rev. C, 38, 1010}

\hi{
Zahn, J.P. 1970, A\&A, 4, 452}

\hi{
\bysame 1977, A\&A, 57, 383}

%\vskip 0.5truecm
\vfill\eject
%%%%%%%%%%%%%%%%%%%% Figures Captions %%%%%%%%%%%%%%
\centerline{\bf FIGURE CAPTIONS}
\vskip 0.3truecm

\noindent
{\bf Figure 1}.---
Dimensionless tidal amplitude associated with a particular g-mode
of coalescing binary neutron stars as a function of the binary separation $D$.
The resonance radius is assumed to be $D_\al/R=8$, and $Rc^2/(GM)=5$,
$q=M'/M=1$, where $R$ is the stellar radius, $M$ is the neutron star mass.
(a) shows the real part of the function $b_\al(t)$ (light solid line)
and $|b_\al(t)|$ (dark solid line) for the $m=2$ mode;
the dashed line corresponds to the
quasi-equilibrium solution (eq.~[3.2]).
(b) shows the real part (solid line) and the imaginary part (dotted line)
of the function $c_\al(t)$ for $m=2$.
The orbital phase function is chosen such that $\Phi(D=0)=0$.

\noindent
{\bf Figure 2}.---
Equations of state for four neutron star models considered in this paper:
(a) pressure $P$; (b) adiabatic sound speed $c_s$ (where $c$ is the speed of
light); (c) the proton fraction $x=n_p/n$;
(d) the fractional difference between
$c_s^2$ and $c_e^2$. The dotted lines are for model AU, the solid lines for
model UU, the short-dashed lines for model UT, and the long-dashed lines for
model UU2. Note that model UU2 has the same pressure and $c_e$ as model UU.

\noindent
{\bf Figure 3}.---
The frequencies of the first ten quadrupole ($l=2$) g-modes based
on four different neutron star models. $n$ specifies the radial
order of the g-mode. The dotted line and round circles are for model AU,
the solid line and squares for model UU, the short-dashed line and triangles
for
model UT, and the long-dashed line for model UU2.

\end